\documentclass[12pt]{iopart}
\usepackage{graphicx,color}
\usepackage{bm}
\usepackage{eucal}
\usepackage{mathrsfs}
\usepackage{subfigure}

\newcommand {\be} {\begin{equation}}
\newcommand {\ee} {\end{equation}}
\newcommand {\Be}{\begin{eqnarray*}}
\newcommand {\Ee} {\end{eqnarray*}}
\newcommand {\bey} {\begin{eqnarray}}
\newcommand {\eey} {\end{eqnarray}}
\newcommand{\bit}{\begin{itemize}}      
\newcommand{\eit}{\end{itemize}}
\newcommand{\bfl}{\begin{flusleft}}
\newcommand{\efl}{\end{flusleft}}
\newcommand{\bfr}{\begin{flushright}}
\newcommand{\bc}{\begin{center}}
\newcommand{\ec}{\end{center}}
\newcommand{\ben}{\begin{enumerate}}    
\newcommand{\een}{\end{enumerate}}

\newcommand{\scs}{\scriptscriptstyle}
\renewcommand{\theparagraph}{\Alph{paragraph}}

\begin{document}

\title[Breathers in proteins]
{Discrete Breathers in a Realistic Coarse-Grained Model of Proteins}


\author{Stefano Luccioli$^{1,2,3}$, Alberto Imparato$^{4}$,Stefano
Lepri$^{1,2}$, Francesco Piazza$^{5}$ and Alessandro Torcini$^{1,2,3,4}$}
\address{$^{1}$ CNR - Consiglio Nazionale delle Ricerche, 
Istituto dei Sistemi Complessi, via Madonna del Piano 10, 
I-50019 Sesto Fiorentino, Italy}
\address{$^{2}$
Centro Interdipartimentale per lo Studio delle Dinamiche 
Complesse, via G. Sansone, 1 - I-50019 Sesto Fiorentino, Italy
}
\address{$^{3}$ Istituto Nazionale di Fisica Nucleare, Sez. Firenze, 
via G. Sansone, 1 - I-50019 
Sesto Fiorentino, Italy}
\address{$^{4}$Dept. of Physics and Astronomy, University of Aarhus,
Ny Munkegade, Building 1520 - DK-8000 Aarhus C, Denmark} 
\address{$^{5}$Universit\'e  d'Orl\'eans and Centre de Biophysique Mol\'eculaire, (CBM-CNRS), 
Rue Charles Sadron, 45071 Orl\'eans Cedex, France}

\begin{abstract}
We report the results of molecular dynamics simulations of an
off-lattice protein model featuring a physical force-field 
and amino-acid sequence. We show that localized modes of nonlinear
origin (discrete breathers) emerge naturally as continuations of a subset
of high-frequency normal modes residing at specific sites
dictated by the native fold. In the case of the small $\beta$-barrel
structure that we consider, localization occurs on the turns connecting 
the strands. At high energies, discrete breathers stabilize the structure by
concentrating  energy on few sites, while their collapse marks the onset of
large-amplitude fluctuations  of the protein. Furthermore, we show how 
breathers develop as energy-accumulating centres following perturbations even
at distant locations, thus mediating efficient and irreversible energy
transfers. Remarkably, due to the presence of angular potentials,  
the  breather induces a local static distortion of
the native fold. Altogether, the combination of this two 
nonlinear effects may provide a ready means for 
remotely controlling local conformational changes in proteins.
\end{abstract}



\submitto{\PB}

\maketitle              

\section{Introduction}
\label{one}

Biopolymers such as proteins and nucleic acids fold into complex three-dimensional structures, whose 
shape is strictly connected to their  biological function~\cite{crei}. The conformation of such 
molecules can change dynamically, in turn modulating the function: for example, activation or 
inactivation of enzymes relies on specific structural modifications occurring at specific 
locations~\cite{Bahar:05,Bahar:99,Eisenmesser:2002vy}.  Typically, such changes are driven 
by either mechanical forces or by converting chemical energy  into conformational 
rearrangements and thus into mechanical work.

Proteins in physiological conditions are immersed in a thermal bath and  
therefore exhibit random thermal fluctuations. 
However, the biological function of a given biopolymer
is often closely  related to a particular kind of motion, typically involving larges-scale 
vibrations~\cite{Westhof:06,Sol:2009zp,Delarue:02,Brooks:05,Elezgaray:02} or the 
fluctuations of entire hinge-domain units~\cite{Brooks:85,Marques:95,Murray:02,Manley:09}.

If collective, low-frequency modes have been traditionally assumed to describe 
functional patterns, there is growing evidence that high-frequency vibrations
contain information on protein stability~\cite{Bahar:1998xu}. Importantly, fast
modes are strongly localized, due to the geometric heterogeneity of  protein
structures. Typically, such vibrations are localized at extremely stiff regions, 
such as hinges, which in turn assume a prominent role in regulating protein
stability and function, as also suggested by experiments~\cite{Eisenmesser:2002vy}.
It is also known that active sites of enzymes have a marked tendency
to  be located within the stiffest segments of the
structures~\cite{Piazza:08,Lavery:07}, which provides a further intriguing
motivation to investigate the connection between localized vibrations and the
peculiarities of protein folds.

When nonlinear effects are considered, the connection between the dynamics of localized modes 
and the details of the scaffolds becomes more interesting. Recently, it has been 
shown that band-edge normal modes (NMs) can be continued to nonlinear localized
vibrations of high energy termed Discrete Breathers (DBs)~\cite{Piazza:08},
also known as intrinsic localized modes~\cite{Sievers:1988or}.
DBs are time-periodic, spatially localized solutions emerging generically in discrete networks of nonlinear 
oscillators~\cite{Flach:1998fj,Flach:2008vx}, that  have been also observed experimentally  in many 
systems~\cite{Flach:2008vx,Trias:2000ej,Binder:2000ih,Sato2009}.

While DBs have been widely studied in spatially homogeneous systems~\cite{Flach:2008vx}, 
the role played by inhomogeneity on their properties remains largely unknown. 
However, the dynamical behaviours of DBs  discovered so far in protein models are remarkable.
In the context of the nonlinear network model (NNM), it has been shown that DBs emerge spontaneously,
upon surface cooling, at few specific sites, invariably within the stiffest
regions~\cite{Juanico:07}. Moreover, they are also able  to self-excite at a target
site upon injecting some energy at a different location, thus mediating  high-yield
energy transfer events~\cite{Piazza:2009jx,Piazza:2009eu,Piazza:2009lq}. In a more
simplified model, it has been shown that DB excitation lowers the free-energy
barrier associated with a given enzyme-catalyzed reaction~\cite{Sitnitsky:2006bq},
thus confirming the  role of protein dynamics in reaction-rate enhancement by
enzymes highlighted by recent  experiments~\cite{Agarwal2005}.

However, DBs in proteins have only been found and characterized so far in extremely simplified
models, either taking into account the three-dimensional folds but with no 
heterogeneity in the force constants~\cite{Piazza:08}, 
or with slightly more elaborate potentials but imposing crudely simplified 
geometries~\cite{Salerno1994263,Archilla:2002bx,SaacutenchezRey2002}. 
Importantly, all studies performed so far notably lacked (i) realistic inter-particle potentials 
and (ii) an explicit account of the amino acid sequence.
In this paper, we make a step forward by investigating DB excitation and their properties 
in a realistic off-lattice model of protein dynamics with coarse-grained 
realistic interaction potentials and a three-code amino-acid sequence. 

The paper is organized as follows. In the next section we introduce our system and describe the  
model, along with an account of our simulation and analysis protocols.
In section~\ref{sec:emergenceDB}, we describe the emergence of DBs as numerical continuation of 
NMs and provide a thorough characterization of their properties.
Finally, we summarize our findings and outline possible future directions prompted by 
our results.

\section{Model and numerical methods}
\label{two}

The 3D off-lattice protein model studied in this paper is a modified version 
of the one initially introduced in Ref.~\cite{honey} and successively 
generalized to include a harmonic interaction between
next-neighbouring beads instead of rigid bonds~\cite{berry}.
This model has been studied to describe thermally-driven
folding and unfolding~\cite{honey,guo_thir,guo,berry,veit,evans,kim,kim-keyes}
and, more recently, to reproduce mechanical manipulation 
experiments~\cite{cinpull,lacks,nostroPRL,nostroPRE,ptp,unftimes}.
It consists of a chain of $L$ point-like
monomers mimicking the residues of a polypeptidic chain
with an associated aminoacid sequence coded by a three-letter alphabet:
hydrophobic ($B$), polar ($P$) and neutral ($N$). 
The intramolecular potential consists of four terms: a stiff
nearest-neighbour harmonic potential, $V_{H}$, intended to maintain the bond
distance almost constant, a three-body interaction $V_{A}$, which accounts for the
energy associated with bond angles, a four-body interaction $V_{D}$ corresponding to
the dihedral angle potential, and a long--range Lennard-Jones (LJ) interaction, $V_{LJ}$,
acting on all pairs $i$, $j$ such that $|i-j| > 2$, namely
\begin{eqnarray}
V_{H} (r_{i,i+1}) &=& \alpha (r_{i,{i+1}}-r_0)^2 
\label{harm}\\
V_{A}(\theta_i) &=& A \cos(\theta_i) +B \cos(2 \theta_i) - V_0
\label{bond}\\
V_{D}(\varphi_i, \theta_i, \theta_{i+1}) &=& C_i [1-
S(\theta_i,\theta_{i+1})\cos(\varphi_i))] 
\nonumber\\
 &+& D_i [1-S(\theta_i,\theta_{i+1})\cos(3 \varphi_i))] 
\label{dih}\\
V_{LJ}(r_{i,j}) &=& \varepsilon_{i,j} \left( \frac{1}{r_{i,j}^{12}} - \frac{c_{i,j}}{r_{i,j}^6} \right)
\label{lj}
\end{eqnarray}
Here, $r_{i,j}$ is the distance between the $i$-th and the $j$-th monomer, while
$\theta_i$ and $\varphi_i$ are the bond and dihedral angles at the $i$-th monomer, 
respectively. 
The parameters $\alpha$ and $r_0=1$ fix 
the strength and the equilibrium distance between consecutive monomers 
along the backbone, respectively.
The term $V_{A} (\theta_i)$ it such that it corresponds, 
up to the second order, to a harmonic  
term $k_\theta (\theta_i -\theta_0)^2/2$, where
\Be
A &=& - k_{\theta} \frac{\cos\theta_0}{\sin^2\theta_0} ,\qquad
B= \frac{k_{\theta}}{ 4 \sin^2\theta_0} ,\qquad
\nonumber \\
V_0 &=& A \cos\theta_0 + B \cos( 2\theta_0) \quad ,
\label{param_harm1}
\Ee
with $k_{\theta} = 20 $, $\theta_0 = 5 \pi/12 \enskip rad$ or $75^{\rm o}$~\cite{karplus}.

The dihedral-angle potential is characterized by three minima for $\varphi = 0$
(associated with the so-called {\em trans} state)  and $\varphi= \pm 2 \pi/3$
(corresponding to {\em cis} states). This term is mainly responsible for the
formation of secondary structures. In particular, large values of the parameters
$C_i, D_i$ favor the formation of the trans state and therefore of
$\beta$-sheets, while when cis states prevail $\alpha$-helices are formed. The
parameters $(C_i,D_i)$ have been chosen as in Ref.~\cite{veit}, {\em i.e.} if
two or more beads among the four defining $\varphi$ are neutral (N) then $C_i =
0$ and $D_i = 0.2$, in all the other cases $C_i = D_i = 1.2$. 
The {\it tapering function} $S(\theta_i,\theta_{i+1})$ has been introduced 
in the expression of $V_{D}$ in order to smooth off the singularities 
introduced by the derivatives of the dihedral angle cosines
(for an exact definition see Refs.~\cite{nostroPRE, rampioni}).

The Lennard-Jones term $V_{LJ}$ describes the interactions with the solvent,
that depend on the nature of the interacting residues as follows:
if any of the two monomers is neutral, the potential is repulsive $c_{N,X} =0$
and its energy scale is fixed by $\varepsilon_{N,X} = 4$;
for interactions between hydrophobic residues $c_{B,B} =1$ and 
$\varepsilon_{B,B} = 4$; for any polar-polar or polar-hydrophobic 
interaction $c_{P,P} \equiv c_{P,B} = -1$
and $\varepsilon_{P,P} \equiv \varepsilon_{P,B} = 8/3$.

According to the above definitions, the Hamiltonian of the system reads
\begin{eqnarray}
H = K + V = \sum_{i=1}^L \frac{|\vec{p}(i)|^{2}}{2} +\sum_{i=1}^{L-1} V_{H}(r_{i,i+1}) +
\nonumber\\
+\sum_{i=2}^{L-1} V_{A}(\theta_i)+ 
\sum_{i=2}^{L-2} V_{D}(\varphi_i,\theta_i,\theta_{i+1}) +
\sum_{i=1}^{L-3} \sum_{j=i+3}^{L}  V_{LJ}(r_{ij})
\label{hamil}
\end{eqnarray}
\noindent
where all monomers are assumed to have the same unitary mass.

\begin{figure}[ht]
\begin{center}
\includegraphics*[width=10.0 truecm,clip]{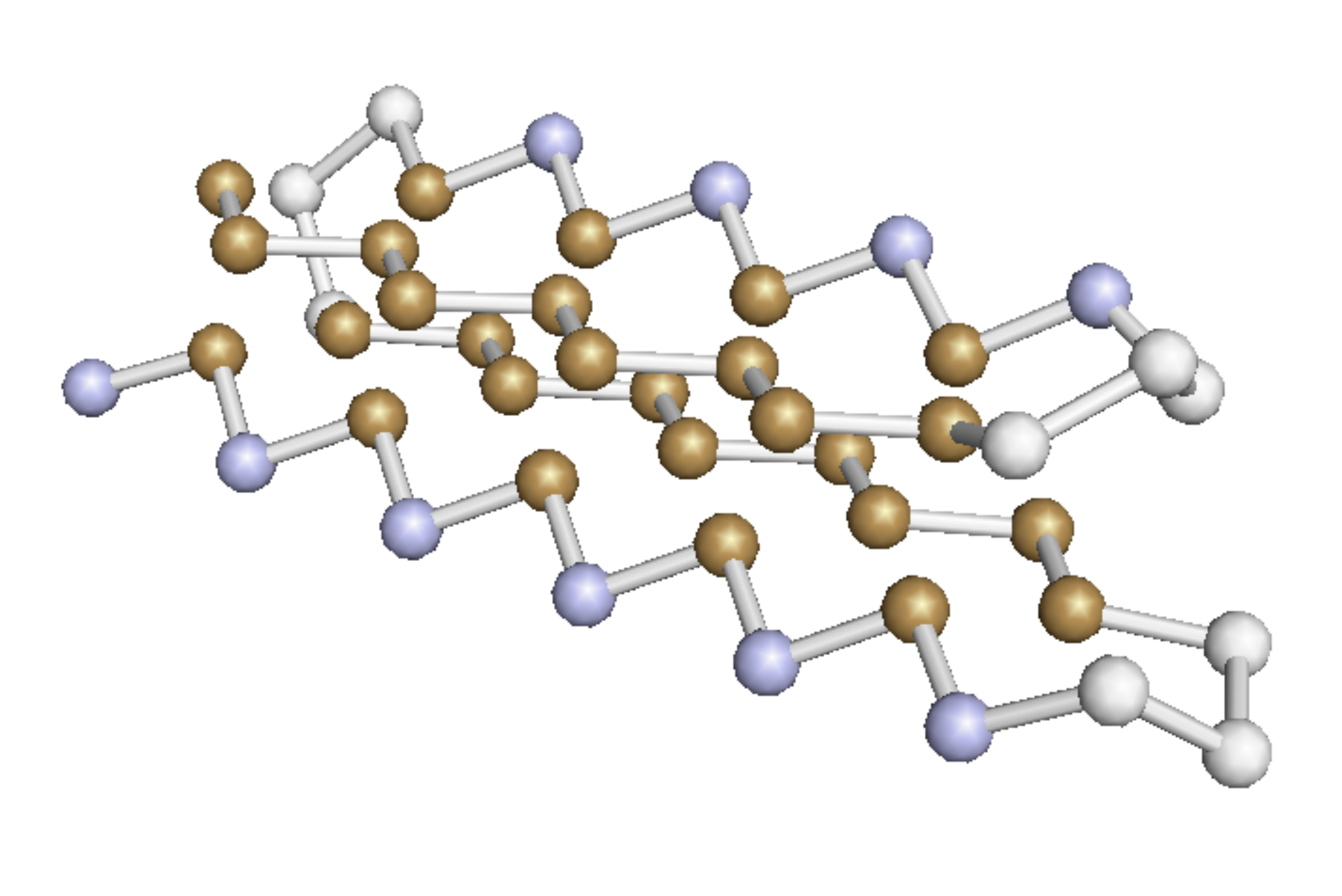}
\end{center}
\caption{(Color online) Native configuration of the protein model. 
Different colors represent the three different types of beads: 
white (neutral), light blue (polar) and sand (hydrophobic). 
Parameters are: $L=46$, $\alpha=1000$.}
\label{nativacolori}
\end{figure}

In the present paper we consider the following sequence of $L=46$ monomers: 
$B_9 N_3 (PB)_4  N_3 B_9  N_3 (PB)_5P$. 
This sequence has been widely analyzed in the past for thermal
folding~\cite{honey,guo_thir,guo,veit,berry,evans,kim,kim-keyes} as well
as for mechanically induced unfolding and 
refolding~\cite{cinpull,lacks,nostroPRL,nostroPRE}. 
Here, we adopt the same potential and 
parameters as in Refs.~\cite{veit,nostroPRL,nostroPRE}, 
except for a stiffer harmonic constant $\alpha$ ($50\le \alpha \le 1000$).
The simulations reported henceforth refer
to $\alpha=1000$, except when otherwise indicated. 
We have verified that this choice does not affect the thermodynamic 
properties of the model ({\em i.e.} the {\em folding temperature} and the 
{\em hydrophobic collapse temperature}) found in Ref.~\cite{nostroPRE} for $\alpha=50$.
This result could be expected, since the harmonic bead-bead interaction
introduced in~\cite{berry} has been already shown not to alter the
folding properties of the original Honeycutt-Thirumalai model~\cite{honey}.

With the above choice of  parameters, the heteropolymer exhibits a 
four-stranded $\beta$-barrel native configuration (NC), described by the 
coordinates $\vec{q}_{NC}(i)$, $i=1\dots L$, and shown in Fig.~\ref{nativacolori}.
The NC corresponds to the absolute minimum of the potential energy,
whose value for $\alpha=1000$ is $V_{\scs NC}\approx -48.94$.
The native structure is stabilized by the attractive hydrophobic interactions among the 
$B$ residues, in particular the first and third $B_9$ strands, forming the core of the NC,
are parallel to each other and anti-parallel to the second and fourth strand
$(PB)_4$ and $(PB)_5P$. The latter are exposed to the solvent 
due to the presence of polar residues. Overall, the four strands are
separated by stretches  of  three consecutive neutral beads, forming as many 
turns (see Fig.~\ref{nativacolori}). These involve the
following sites: 10-12 (first turn), 21-23 (second turn) and 
33-35 (third turn).

In the following, we report equilibrium microcanonical
results obtained by integrating  Hamilton's equations 
by means of a fourth order symplectic integrator~\cite{McLachlan} 
with a time-step of $10^{-3}$ time units  (ensuring a 
relative energy conservation of $\sim 10^{-7}$). 
In all cases, the initial coordinates of the beads were taken
to correspond to the NC (zero-displacement).

\subsection{Normal modes analysis}
\label{three}

The properties of nonlinear modes in spatially and force-heterogeneous systems 
strongly depend on the features of the linear spectrum:
in particular, non-resonant nonlinear modes may emerge within inter-mode gaps
and highly localized vibrations can be associated with band edges~\cite{Piazza:08}. 
Hence, a detailed  understanding of the linear spectrum is an
essential pre-requisite for our analysis. More precisely, we wish to investigate
how the normal mode frequencies and eigenvectors depend on the harmonic
bond stiffness $\alpha$. 

\begin{figure}[ht]
\begin{center} 
\includegraphics[scale=0.4,clip]{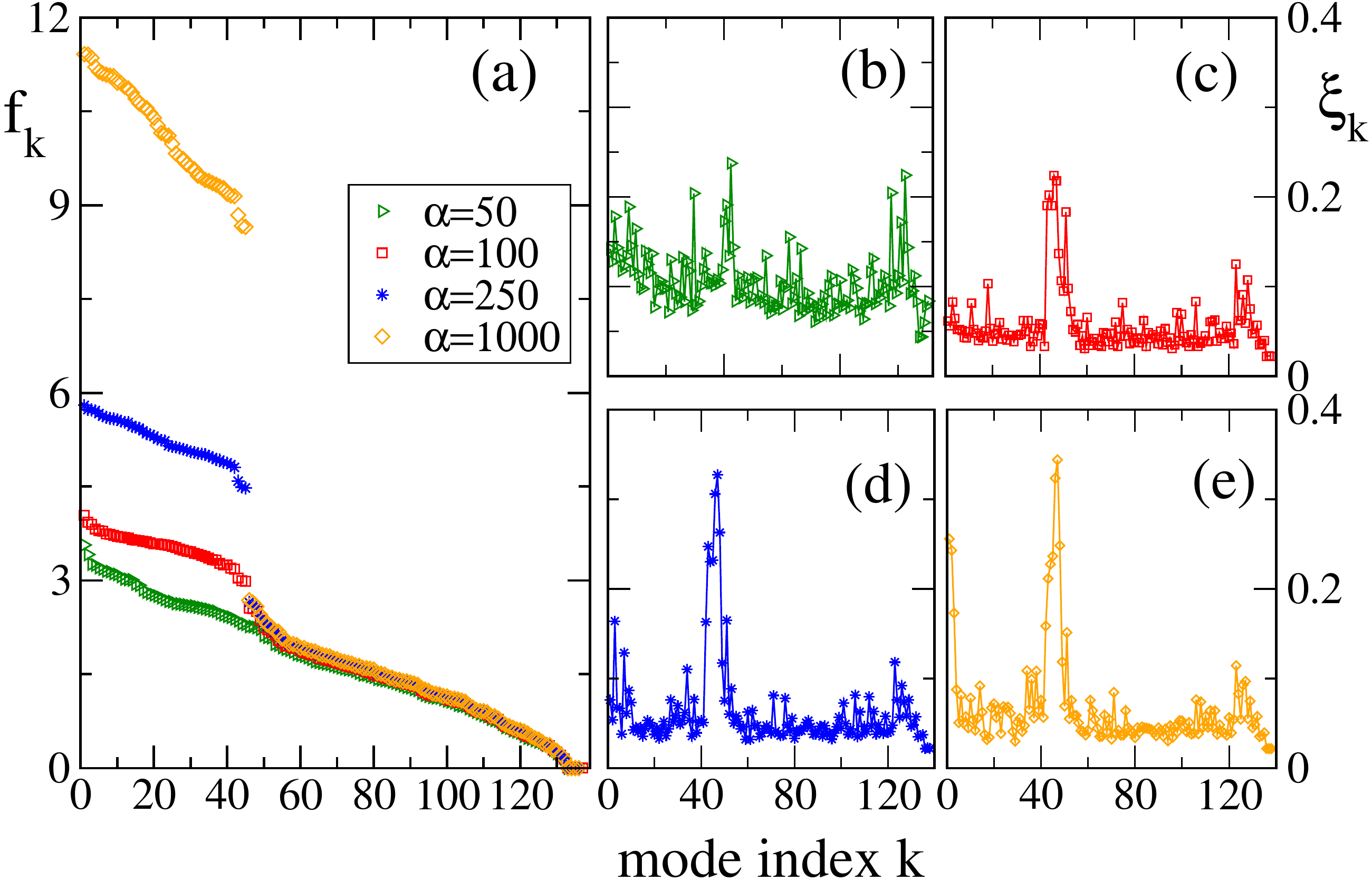}
\end{center}
\caption{(Color Online) (a) Frequencies of the NMs, $f_{k}$. (b),(c),(d),(e) 
Inverse of the participation ratio for the corresponding 
eigenvectors for increasing value of the stiffness of the harmonic bonds. 
Triangles, squares, stars, and circles refer 
to $\alpha=50,100,250$, and $1000$, respectively.}
\label{figuravarioalpha}
\end{figure}

\begin{figure}[ht]
\begin{center} 
\includegraphics[width=0.6\textwidth,clip]{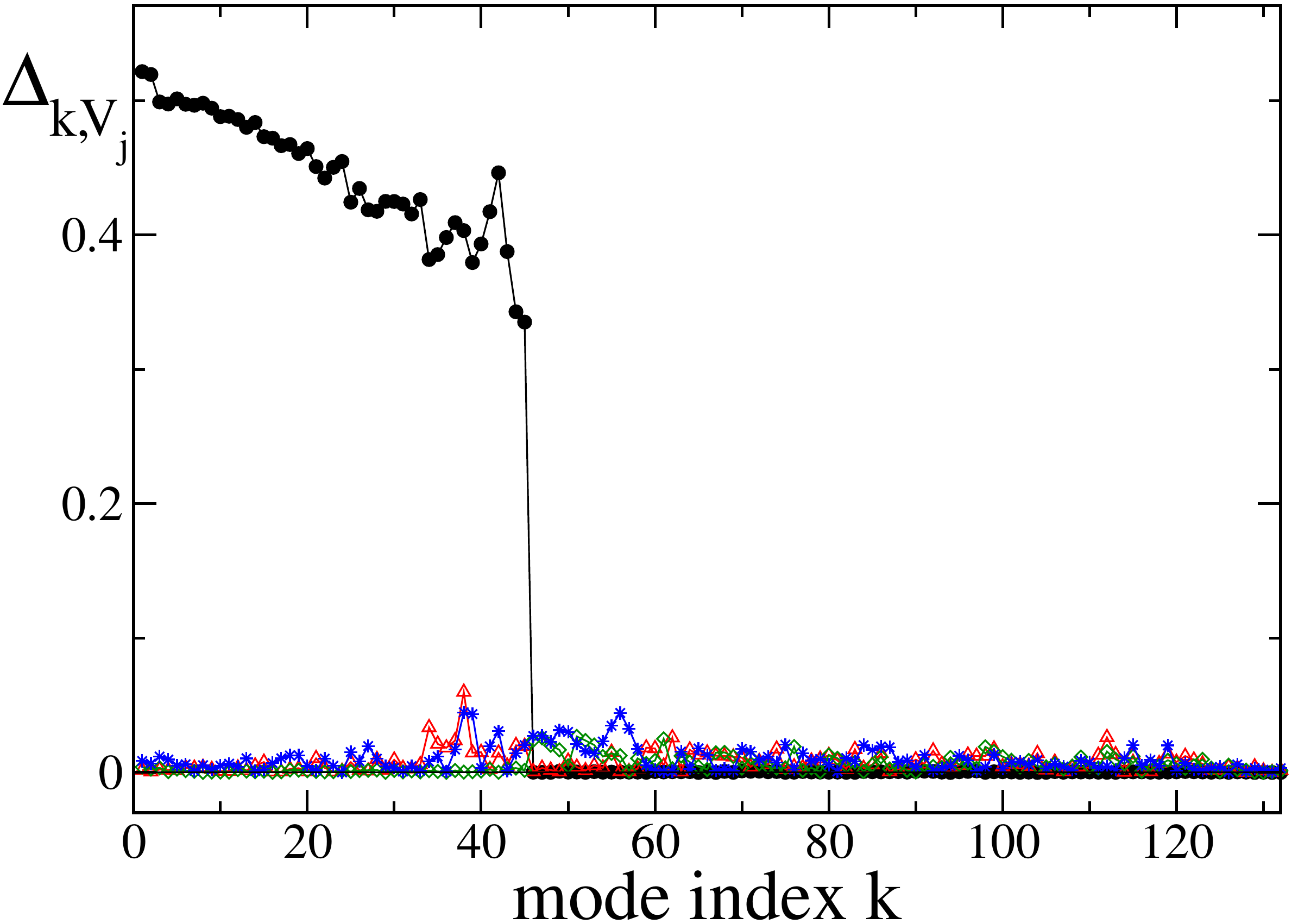}
\end{center}
\caption{(Color online)  Relative variations of the energetic contributions 
due to a perturbation of the NC along 
the direction of the  $k$-th NM, as given by eq.~(\ref{kappaV}), for $\alpha=1000$.
Circles, triangles, diamond and  stars 
correspond to $\Delta_{k,V_{H}}$, $\Delta_{k,V_{A}}$, $\Delta_{k,V_{D}}$, 
$\Delta_{k,V_{LJ}}$, respectively. 
}
\label{spostamenti}
\end{figure}

The NM spectrum associated with the NC of our model protein is composed by
$3L-6$ non-zero frequencies $\{f_{k}=\sqrt{\lambda_{k}}/2 \pi
\}$, $\lambda_k$, being the eigenvalues of the Hessian matrix~\cite{wales}. The NM spectra 
are reported in Fig.~\ref{figuravarioalpha} (a) for values of $\alpha$ ranging from 50 to 1000. 
The most important observation for our purposes is that, 
for large enough values of $\alpha$, the spectrum splits in two bands separated by a gap.
The highest portion (the optical band) comprises $L-1$ modes associated  with
the stiffest force constants, that is 
bond (backbone) distortions along the chain, while the lowest part (the acoustic
band) is composed by the remaining $2L-5$ modes that represent collective
motions of the beads which are hardly affected  by changes in $\alpha$.  
This interpretation can be substantiated by the following analysis.
We have first slightly perturbed 
the NC along the direction of each eigenvector with an arbitrary
amplitude $\varepsilon$, {\em i.e.}
$\vec{q}_{NC}(i)\rightarrow \vec{q}_{k}(i)=\vec{q}_{NC}(i)+\varepsilon \, \vec{e}_{\, \, k}(i)$. 
Then we have evaluated the relative variations of the potential energy
components $V_{j}$, $j=H$ (harmonic), $A$ (three-body angular), 
$D$ (diehdral), $LJ$ (Lennard-Jones), with respect to their
values in the NC according to the following definition:  
\begin{equation}
\Delta_{k,V_{j}}=\frac{|V_j(\vec{q}_k)-V_j(\vec{q}_{NC})|}{|V_j(\vec{q}_k)|} 
\quad ,
\label{kappaV}
\end{equation}
As it is clear from Fig.~\ref{spostamenti}, the
perturbation along the first 45 eigenvectors is only restricted 
to the degrees of freedom associated with bond deformation, {\em i.e.} to the 
harmonic contribution to the potential energy.

Furthermore, the formation of the gap is accompanied by a localization of the eigenvectors
around the gap-facing band edges. In order to characterize the degree 
of localization of the $k$-th
NM $\vec{e}_{\, \,  k}(i)$ ($i=1,2,\dots,L$), we measured
its {\it inverse participation ratio}~\cite{aoki},
\begin{equation}
\xi_{k}=\sum_{i=1}^L |\vec{e}_{\, \,  k}(i)|^4
\end{equation}
where the eigenvectors are normalized to unity. 
For an eigenvector localized on a
single site $\xi \simeq 1$, while for a completely delocalized 
state $\xi \simeq 1/L$. Therefore, the more the eigenvector is localized
the larger is $\xi$. The inverse participation ratio is plotted
in Fig.~\ref{figuravarioalpha} (b-e) for different  values of $\alpha$.
In particular, for $\alpha=50$ all the eigenvectors are
extended (see Fig.~\ref{figuravarioalpha} (b)). However, by increasing $\alpha$, 
the degree of localization of NMs in the lower optical band edge
($k=43,44,45$) and of the first three NMs of the acoustic band edge
($k=46,47,49$) increases (Fig.~\ref{figuravarioalpha} (c-e)). A 
further increase in stiffness does
not only lead to an enhancement of localization in the proximity of the gap, but
also determines the localization of the first three optical eigenvectors
($k=1,2,3$), see Fig.~\ref{figuravarioalpha} (e). It is
also interesting to observe that for increasing $\alpha$ the first and last
three optical frequencies  detach more and more from the band core (see
again Fig.~\ref{figuravarioalpha} (a)).
In the following, we will fix $\alpha=1000$.

\begin{figure}[ht]
\begin{center} 
\includegraphics[scale=0.4,clip]{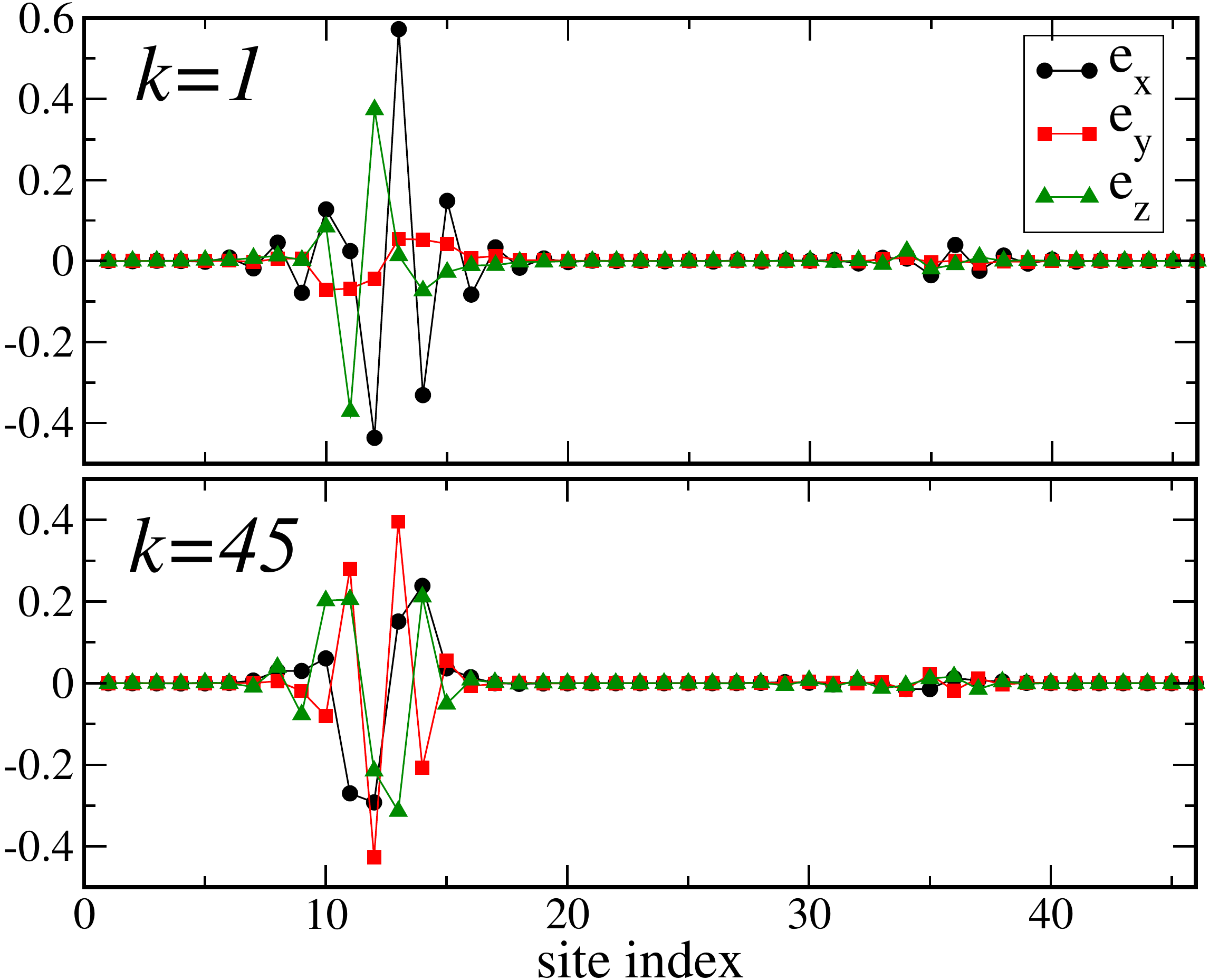}
\end{center}
\caption{(Color Online) Cartesian components of the eigenvectors $\vec{e}_{\, \, 1}$ (top)
and $\vec{e}_{\, \, 45}$ (bottom) corresponding to the frequencies at the 
upper and lower edges of the optical band, respectively, for $\alpha=1000$.
}
\label{modinormali}
\end{figure}

We notice that the most localized modes of the optical band
are localized at the three turns of the NC. 
More precisely, modes $k=1,45$ at the first turn (Fig.~\ref{modinormali}), 
NMs $k=2,44$ at the third turn and modes $k=3,43$ at the second one.
The modes at both edges of the optical band are characterized by single or 
groups of neighbouring oscillators in opposition of phase,  as shown in
Fig.~\ref{modinormali}.  Qualitatively, one may interpret them as impurity modes 
with the turns acting as structural defects~\cite{impurity}.
Interestingly, for a toy model reproducing a single bent 2$D$ chain with fixed curvature,
Archilla {\it et al}~\cite{Archilla:2002bx} also reported a pair of localized 
modes lying at the  band edge, with
frequencies slightly detached from the band itself.
Concerning the acoustic band-edge modes $k=46,47,48$,
only the $y$-component appears localized on the first and third $\beta$-strand $B_9$,
with the oscillators on the first and third strands in opposition of phase.
Since these strands represent the core of the protein, we expect
that excitation of these modes should cause large rearrangements in the structure, 
leading to protein unfolding.

\section{Emergence of discrete breathers}
\label{sec:emergenceDB}

In this Section we show how DBs can be created by exciting the 
NC along the direction of certain NMs. 
In these simulations, we initialize the beads positions in the NC
and assign the initial velocities proportional to the pattern of 
the selected NM. The amplitudes of the kinetic energy perturbation 
$K_0$ will be henceforth measured in temperature units $T=2K_0/3L$, 
where a unitary Boltzmann constant is assumed.

As we shall show, the excitation of a normal mode provides an
effective means of feeding energy to a discrete breather. However, due to specific 
selection rules matching spatial overlap between normal modes~\cite{Kidera:00}, 
a portion of the initial energy necessarily flows into a number of other modes.
Such {\em background radiation} competes with the nonlinear localization process. This in turn, as we 
shall see in the following, accelerates the collapse of nonlinear modes.
In order to analyze DB properties, it proves
useful to get rid of background radiation through {\em surface damping}~\cite{Juanico:07}.
To this end, in the first part of the work we shall cool down the velocities of the
beads at the chain terminals. 
More precisely, we integrate Hamilton's equations of motion while rescaling, 
at regular time intervals $\Delta t$,  the momenta 
of the beads located at the  chain terminals 
$i= 1, \dots, 4$ and $i=43,\dots,46$), that is
\begin{equation}
\vec{p}(i) \rightarrow \vec{p}^{\enskip \prime}(i)=(1-\gamma) \vec{p}(i)   
\hspace{0.5cm} {\rm with} \hspace{0.5cm} \gamma \le 1 \quad .
\end{equation}
Unless otherwise specified, the cooling parameters are fixed to 
$\gamma=0.001$ and $\Delta t=0.02$. In section~\ref{nocooling} a comparison 
between the dynamics with and without  cooling  is presented.

\begin{figure}[ht]
\begin{center} 
\subfigure[]{
\includegraphics[width=0.5\textwidth,clip]{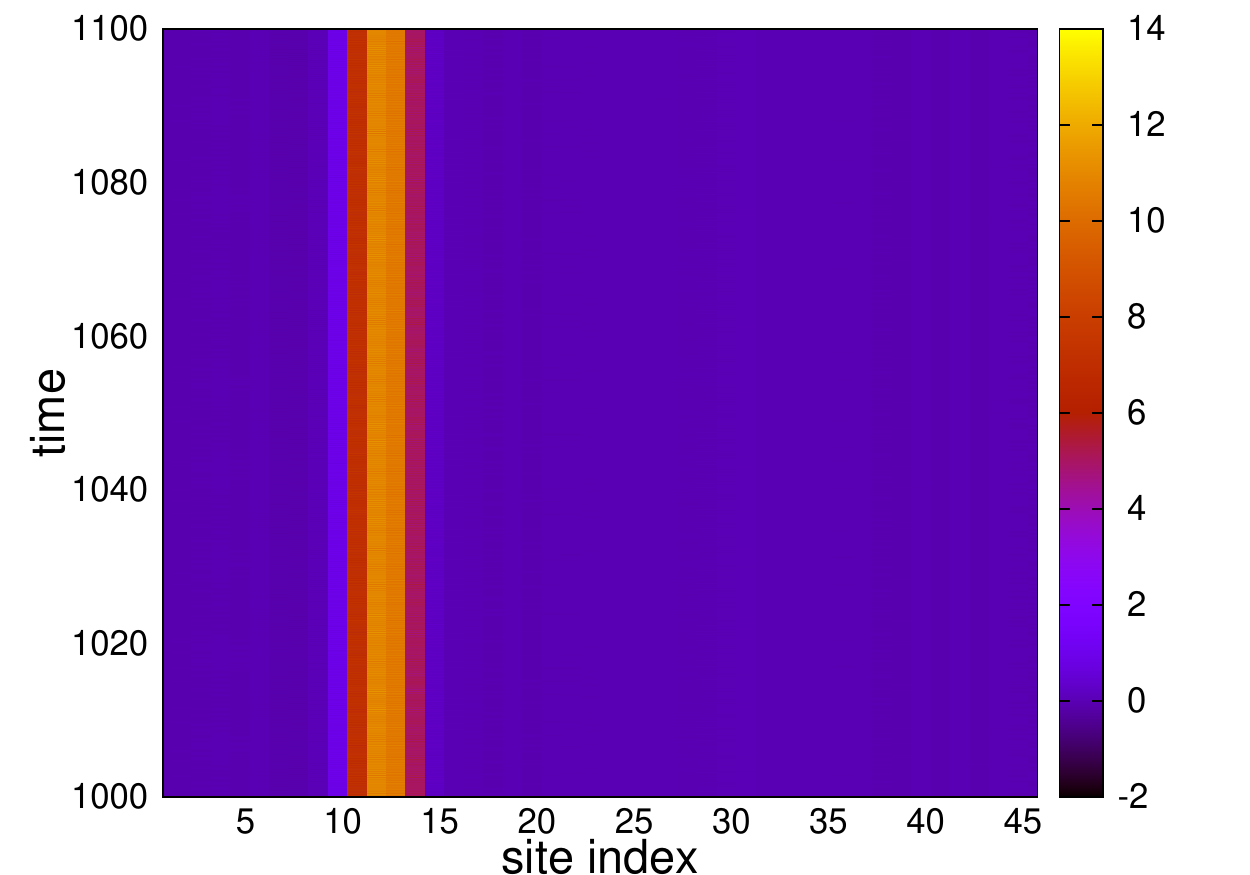}}
\subfigure[]{
\includegraphics[width=0.4\textwidth,clip]{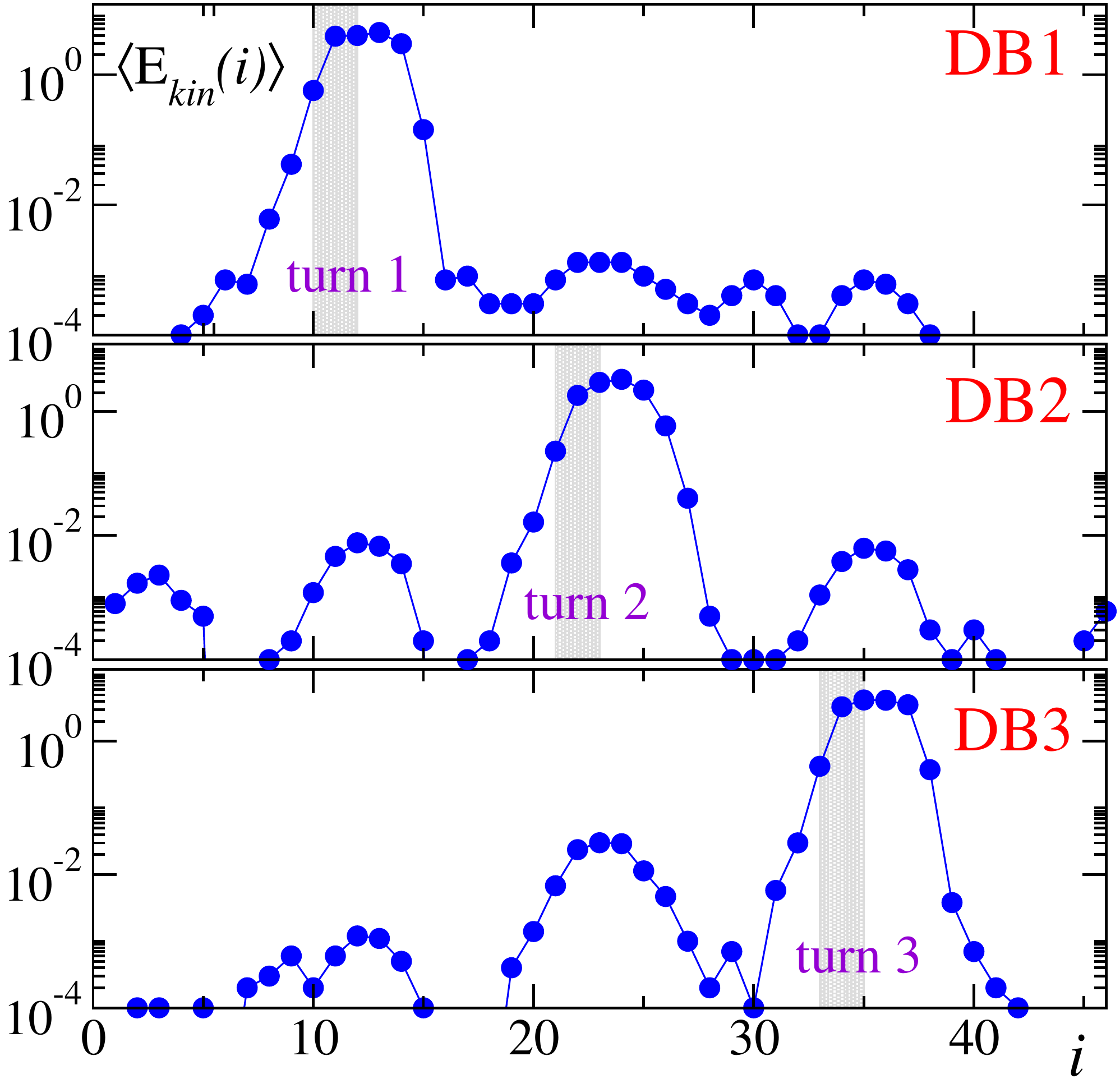}}
\end{center}
\caption{(Color Online) (a) Total energy difference per site
$\Delta E_{tot}(i)$ with $i=1, \dots,L$, as a function of time.
for DB1 with an initial temperature $T=0.63$. (b) Kinetic site energy $\langle E_{kin}(i) \rangle$ averaged
over a time span of the order of $50-100$ breather's periods
in order to smooth out fluctuations. From top to bottom
DB1 with initial temperature $T=0.63$, DB2 at $T=0.42$ and DB3 at $T=0.63$.
Cooling parameters are $\gamma=0.001$, $\Delta t=0.02$ (a) 
and $\Delta t=0.005$ (b).  Grey regions mark the protein turns (see text).
}
\label{modo45_t0.3} 
\end{figure}

In order to visualize the energy localization along the chain  highlighting the
emergence of a DB, we recorded the site kinetic energy $E_{kin}(i)$ and the
total excitation energy per site $\Delta E_{tot}(i)=E_{kin}(i)+V(i)-V_{\scs NC}(i)$,
where $V(i)$ is the potential energy contribution which can be associated with
the $i$-th bead (see  Appendix A for the exact definition) and $V_{\scs NC}(i)$ is
the corresponding value in the NC. The presence of a DB is detected as a
protracted  concentration of energy on a limited  number of sites.  

Our first important observation is that a stable DB can be created 
by perturbing the NC along the direction of \textit{each of the first
45 modes}, or along a linear combination of them.
An example is reported in Fig.~\ref{modo45_t0.3} (a) for the excitation 
of the lower optical-band edge mode. 
Remarkably, only three distinct breathers were observed to emerge despite the
different directions of the employed perturbations. They are
located close to the three turns and we thus 
denote them as DB1, DB2, and DB3, after the index of the corresponding
turn (Fig.~\ref{modo45_t0.3} (b)). 

A peculiarity of such localized modes is that  
they feature patterns that are not simply localized over a few adjacent sites. 
Rather,  they display non-negligible components localized elsewhere in the
chain. This can be appreciated by looking to the average site kinetic energy
reported  in Fig.~\ref{modo45_t0.3} (b) for the three breathers. The largest
contribution to the DB total energy  comes indeed   from the few sites located
at the corresponding turn. However,  energy components two-orders of magnitude
smaller can be found over  the other turns too. The origin of this peculiar
localization patterns can be found in the  three-dimensional structure of the
protein, and it is induced by the requirement of momentum conservation. In other words, the
special built-in momentum-conservation  properties of NM patterns are reproduced
by DBs. As a matter of fact, it turns out that  a mode that is localized at a
few sites conserves its momentum by fractioning  an equal and
opposite amount of momentum among few other locations, instead of  spreading it
over the whole structure. The interesting question arises whether this is a
generic feature of NMs in proteins or it is rather the expression of the peculiar 
native fold considered here. 

\begin{figure} 
\begin{center} 
\includegraphics[width=0.5\textwidth,clip]{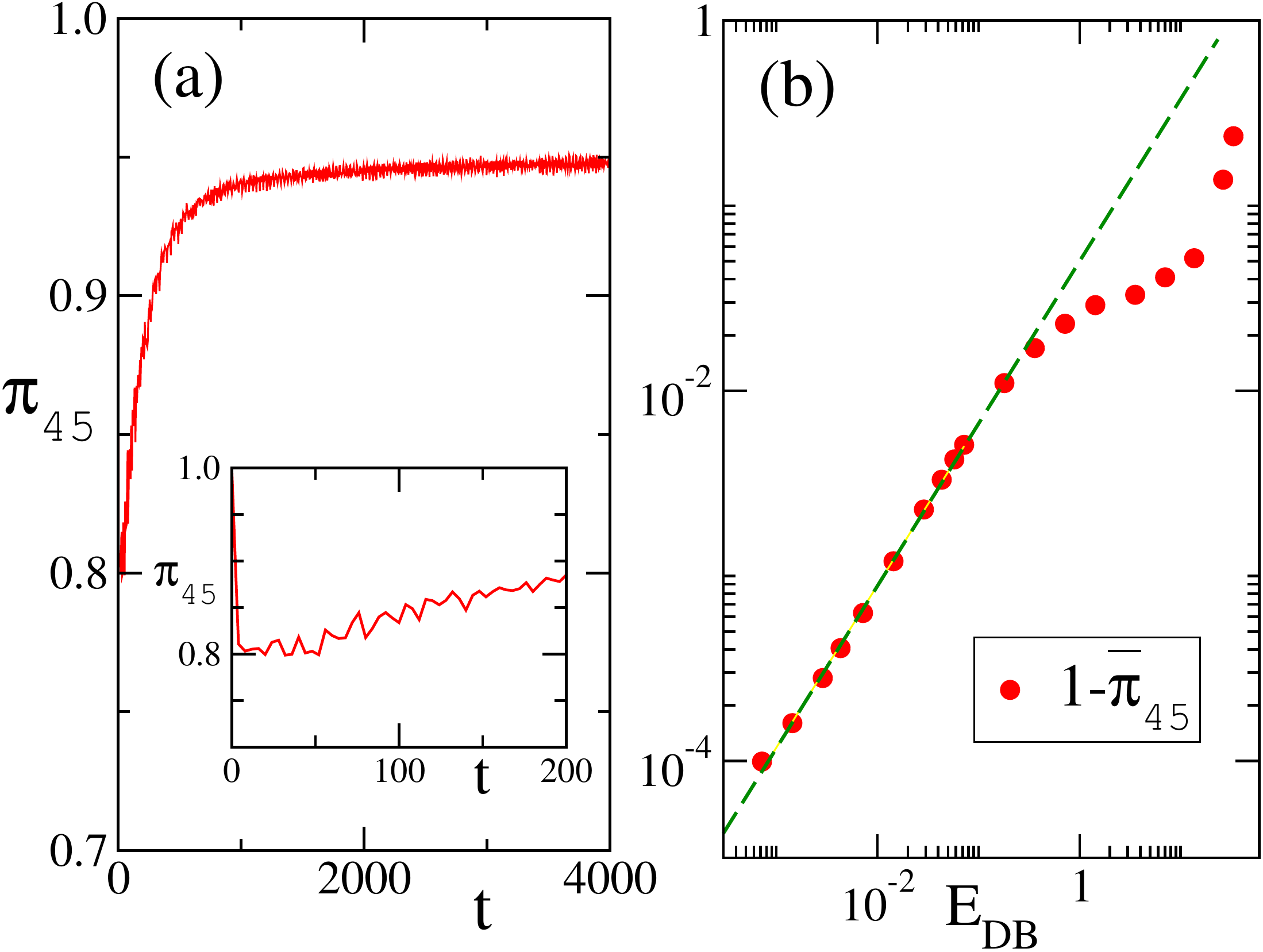}
\end{center}
\caption{(Color online) DB obtained as a continuation of NM $k=$45. (a) Time evolution 
of the projection $\pi_{45}$ for $T=0.21$. The inset shows a close-up of the short-time
dynamics. (b) $1-\overline{\pi}_{45}$ as a 
function of the total energy $E_{\scs DB}$. The dashed line is a power 
law of the type $1 - \overline{\pi}_{45} \propto E_{\scs DB}^\eta$ with $\eta \approx 0.88$.  }
\label{f:proj_DB_modes_temp}
\end{figure}

The three DBs can be considered as nonlinear continuations of  three
corresponding NMs, that are localized at the same turns and whose vibration
patterns can be considered as precursors of the DB displacement fields. 
This is confirmed by looking at the normalized projections $\pi_{k}$ 
of the breathers' velocity fields on  the corresponding modes, defined as: 
\begin{equation}
\pi_{k}= \left\langle \sum_{i=1}^{L} \frac{\vec{p}(i)}{\sqrt{2K}} \cdot \vec{e}_{k}(i) \right\rangle_{t}
\label{def_proiezioni}
\end{equation} 
where the angular brackets denote the average over a window 
of $w$ time units ($w=4.0$ in the following). 
Fig.~\ref{f:proj_DB_modes_temp}~(a) shows that, for DBs originating 
from NM 45,  after an initial transient $\pi_{45}$ 
shows a stable trend with very small residual oscillations around the mean value 
$\overline{\pi}_{45}$ evaluated in the last stage of the simulation ($3000\le {\rm time} \le4000$). 
The projections on the other modes (not shown in the figure) 
are smaller than 0.015. In Fig.~\ref{f:proj_DB_modes_temp}~(b) we plot
$1-\overline{\pi}_{45}$ as a function of the DB total energy, $E_{\scs DB}=K+V-V_{\scs NC}$, 
measured in the last stage. The figure shows that 
$1-\overline{\pi}_{45}$ vanishes when the energy is decreased following a 
nearly linear trend. 

These results confirm and generalize what reported in the context 
of the Nonlinear Network Model (NNM), where gap-less breathers were 
shown to arise as continuations of edge NMs~\cite{Piazza:08}. Also 
in the framework of the NNM, few special regions were shown to act as 
energy-accumulating centres upon generic excitation of the system, 
exactly as we observed here~\cite{Juanico:07}. 

\begin{figure}[ht]
\begin{center} 
\subfigure[]{
\includegraphics[width=9truecm,clip]{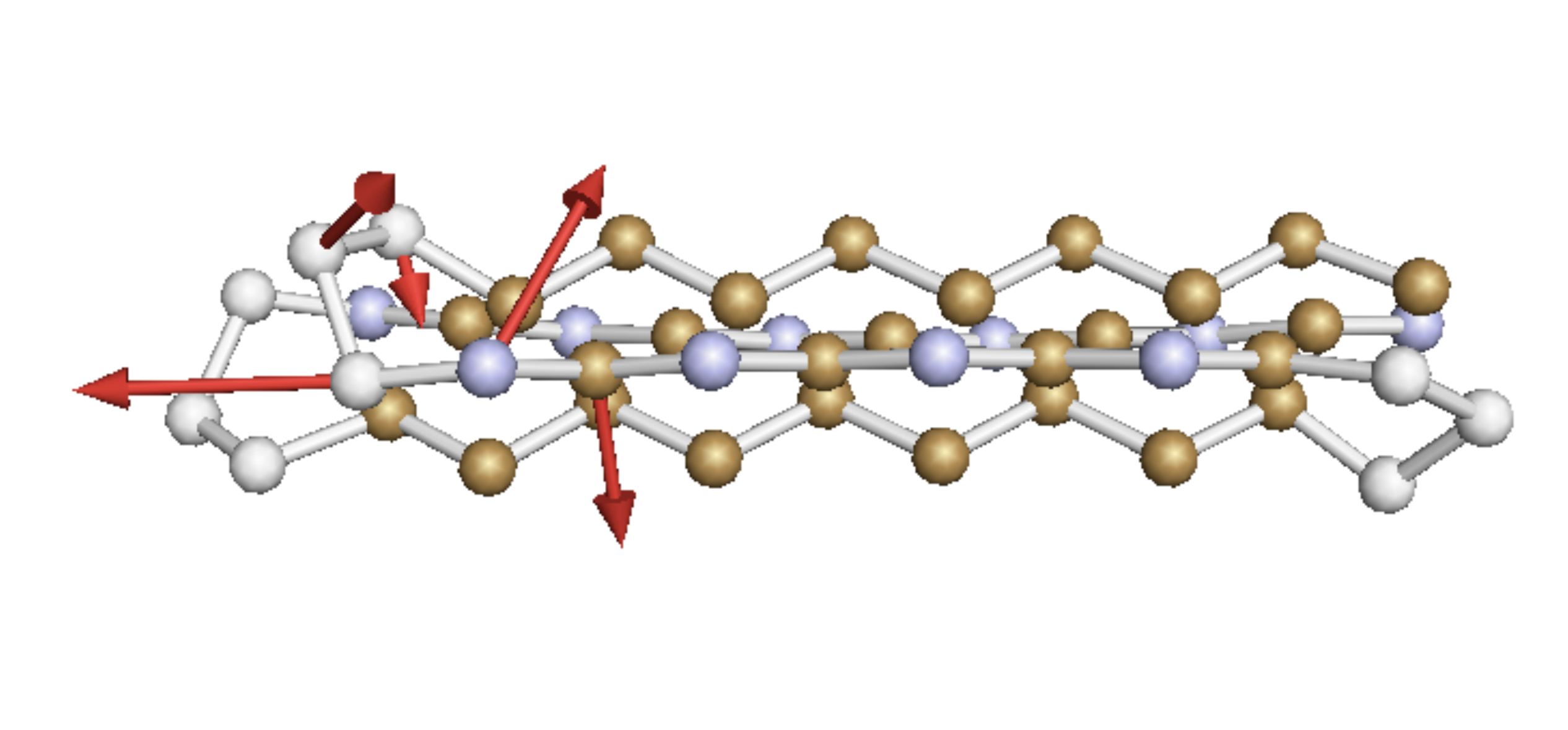}}
\subfigure[]{
\includegraphics[width=5.5truecm,clip]{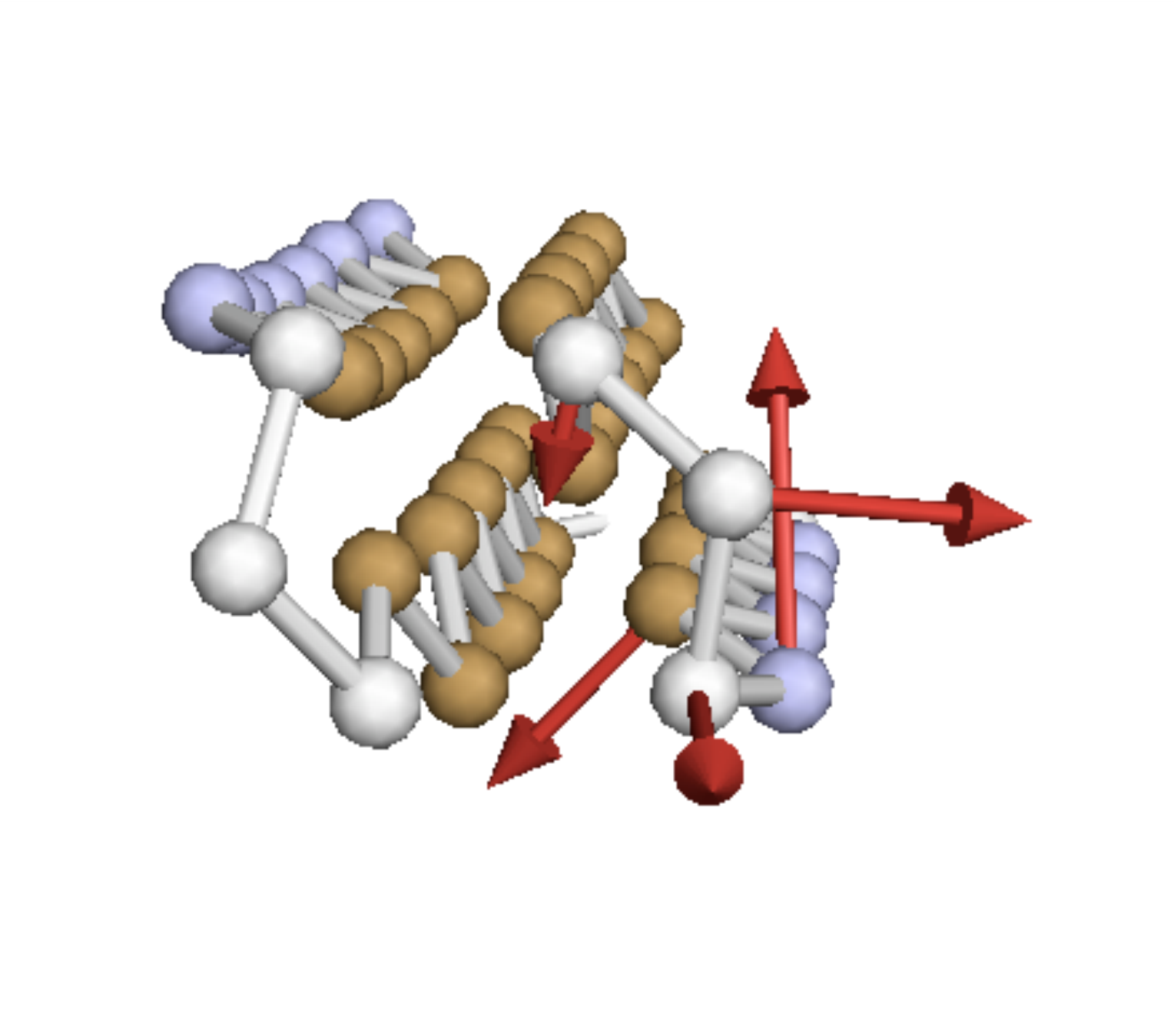}}
\end{center}
\caption{(Color online)
Displacement field of DB1. Side (a) and front (b) views 
(only displacements larger than 1.0 have been 
shown as arrows). The initial condition corresponds to $T=0.63$.
}
\label{figura3d}
\end{figure}

Perturbations along all the optical modes of amplitude $T=0.63$ resulted in the
excitation of DB1 or DB3, while DB2 was observed only in a few cases. The reason
why DB2 seems somehow more difficult to excite is likely to depend upon the
peculiar structural neighborhood of the second turn. This region lies deeper
inside the core of the protein, where beads are more constrained, thus hindering
the DB oscillations (see also the cartoon  showing DB patterns in Fig.~\ref{figura3d}).

%
\begin{figure}[ht]
\begin{center} 
\includegraphics[width=10truecm,clip]{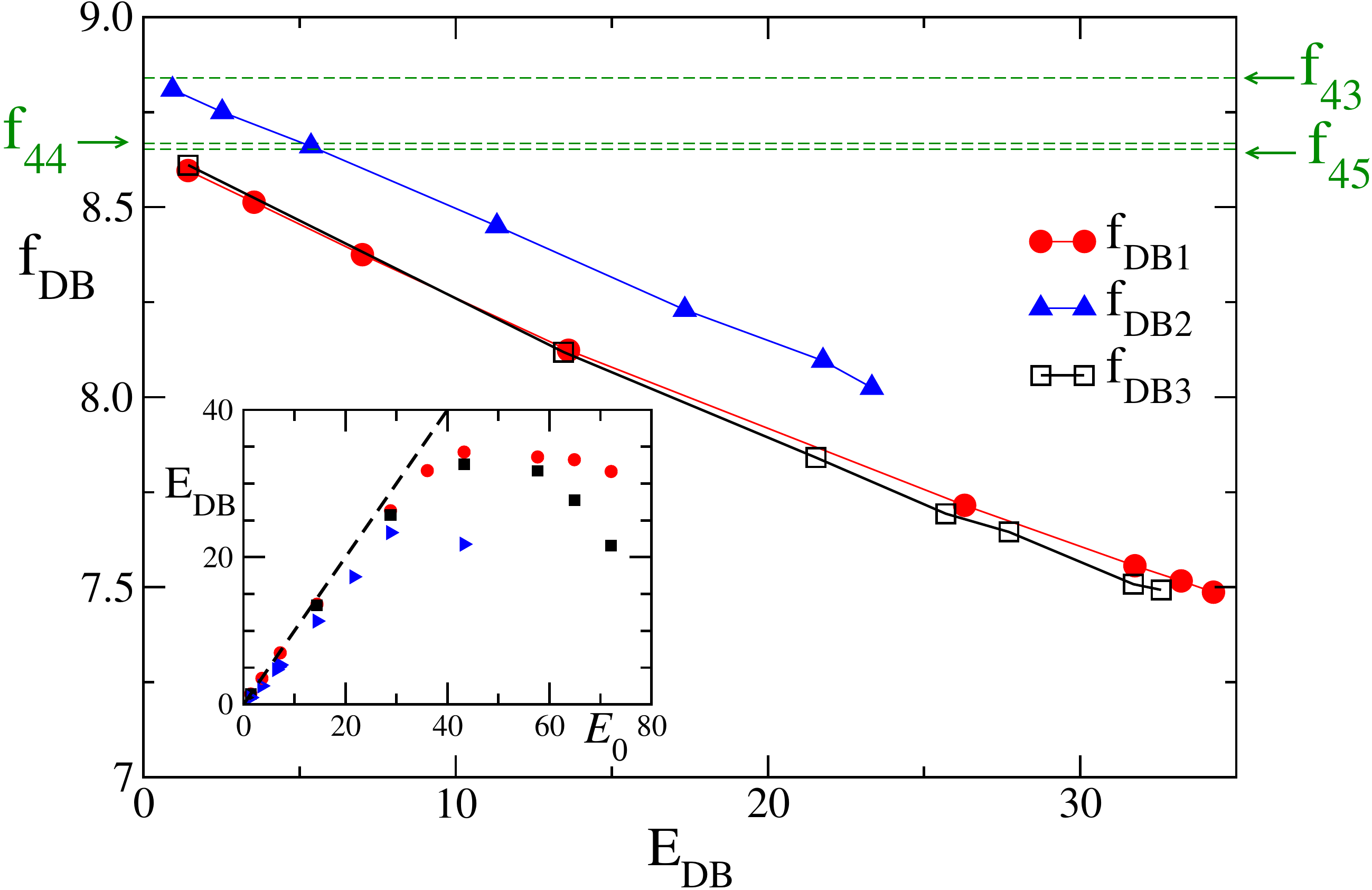}
\end{center}
\caption{(Color online)
DBs originating from $k=$45 (DB1), $k=$43 (DB2) 
and $k=$44 (DB3). Frequencies of DBs as a function of their total energy $E_{\scs DB}$. 
The horizontal dashed lines flag the frequencies of the corresponding NMs.
The inset displays $E_{\scs DB}$ as a function of the initial energy $E_0=K_0-V_{\scs NC}$
for the three DBs: circles, triangles, squares refer to DB1, DB2, DB3, respectively.  
The dashed line marks the {\em full-efficiency} regime $E_{\scs DB} = E_0$.   
}
\label{frequenze_DB}
\end{figure}
%

The DB frequencies $f_{DB1}$, $f_{DB2}$  and $f_{DB3}$  are computed from the
power spectrum of the displacements of the most energetic bead within each DB.
Typical DB spectra feature a main peak and minor (several orders of magnitude 
smaller) peaks signalling the residual excitation  of few normal modes.
The measured frequencies are shown in Fig.~\ref{frequenze_DB} as functions 
of the DB total energy $E_{DB}$, all three DBs exhibit an almost linear decrease
characterized by the same  dispersion relation
$f_{\scs DB}(E_{\scs DB}) = f_{k}( 1 - E_{\scs DB}/\varepsilon)$,
where $f_k$ is the frequency of the NM from which the DB originates and 
$\varepsilon \approx 250$. 
We notice that $\varepsilon$
could be in principle calculated through  Lindstedt-Poincar\'e perturbation theory 
as done, for example, in Ref.~\cite{Piazza:08}. 

The decrease of the DB frequency with the energy $E_{DB}$ suggests that
the relevant nonlinearity is of the \textit{soft} type~\cite{Flach:1998fj}.
This explains why the presence of a gap is necessary for the
formation of localized nonlinear modes. Indeed, we were  unable to excite DBs 
in cases where the spectrum is gap-less.

The frequencies $f_{DB1}$ and $f_{DB3}$ originate from the bottom of 
the optical band ($f_{44}\approx f_{45}$) and  enter the 
gap for arbitrary small energies.
On the contrary, $f_{DB2}$ lies between $f_{43}$ and $f_{44}$ for 
small energies, staying manifestly non-resonant with
the neighbouring optical modes. This behaviour is reminiscent of intra-band DBs 
predicted by Kopidakis and Aubry in nonlinear disordered chains 
~\cite{Kopidakis:2000lr,Kopidakis:1999fk} and  found 
analytically in nonlinear network models of proteins~\cite{Piazza:08}. 
For larger energies, the frequency decreases further and also these DBs eventually
enter the gap. 

It is instructive to consider whether a given DB can be created with 
arbitrary energy by exciting a given normal mode. The inset in Fig.~\ref{frequenze_DB} shows the 
relation between the initial total 
energy $E_0 = K_0 - V_{\scs NC}$ and the energy that is found stored in the DB.
The ratio of the two values can be interpreted as an estimate of {\em transfer efficiency}, 
i. e. the fraction of the initial excitation energy channeled into the DB
and consequently pinned down at a very specific location.
Remarkably, we see that all the three DBs are characterized by a nearly unitary 
efficiency, up to a point where the curve saturates and then starts bending down. 
The transfer of energy from the NM to the DB is nearly optimal up to initial energies of 
the order $E_0 = K_0 - V_{\scs NC} \approx 30$. This indicates the existence of a sort of 
maximal "DB capacity'' -- if the NM is feed with larger energies, the excess energy 
cannot be further  injected in a localized mode and it is necessarily spread 
across the whole structure.

Finally, no DB  could be excited by perturbing  the structure along acoustic
NMs. This is most likely a consequence of the soft nonlinearity.
Moreover, we were also unable to excite DBs through  arbitrary localized
perturbations of the NC.  Indeed, initial conditions of the latter type have non-zero
components over many NMs,  including the acoustic ones.

%
\subsection{DBs originating from the lower optical-band edge}

\begin{figure}[ht]
\begin{center} 
\includegraphics[width=0.6\textwidth,clip]{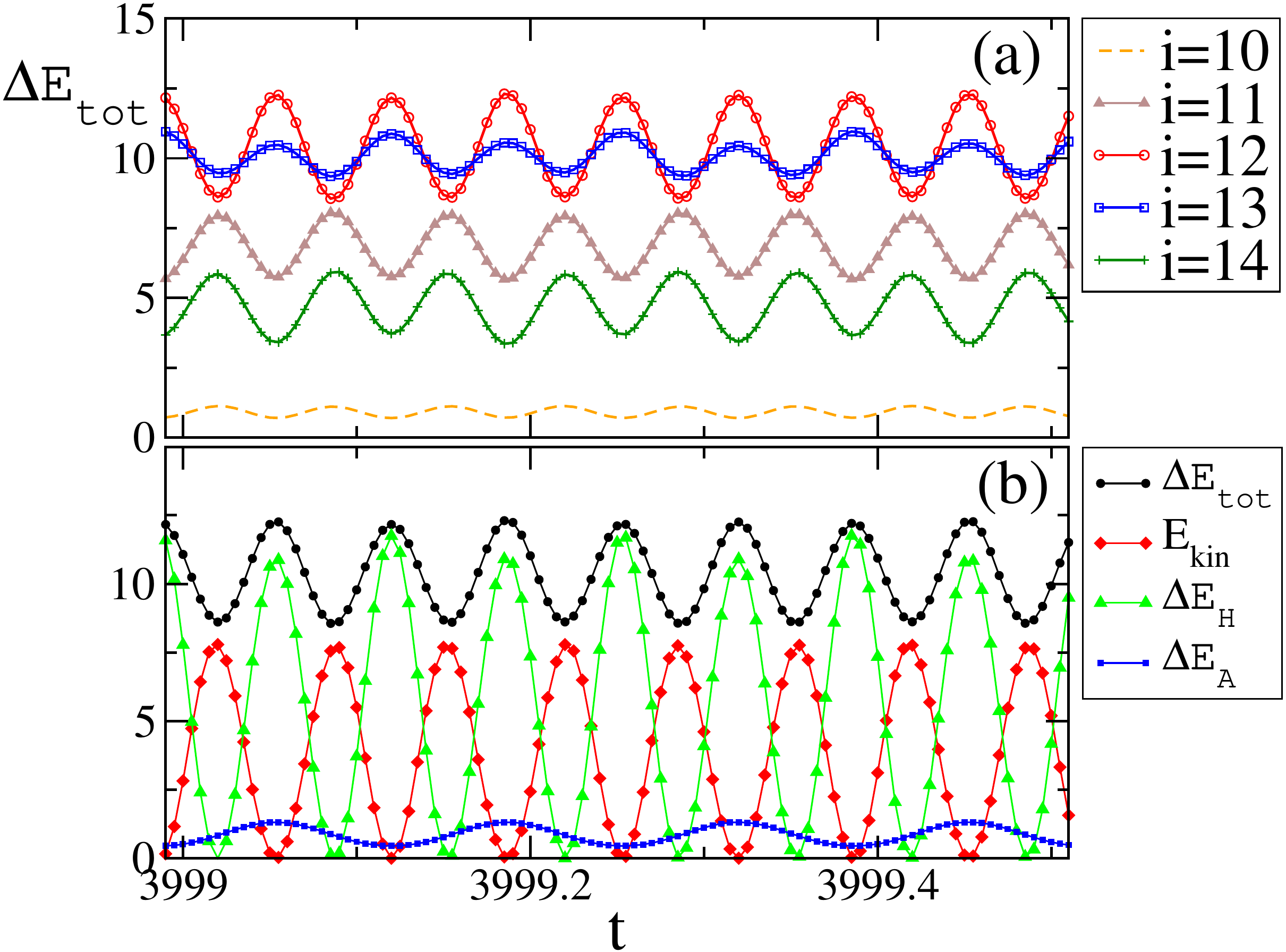}
\end{center}
\caption{(Color online) (a) Total site energies for DB1 and (b) 
the different energy contributions at site $i=12$ versus time 
($\Delta E_{j}=V_{j}-V_{\scs NC}(j)$ where $j=H,A$, the dihedral and Lennard-Jones 
contributions being negligible).
Parameters are: $T=0.63$, $\gamma=0.001$, $\Delta t=0.005$.}
\label{energiaDBmodo45}
\end{figure}

In this section we shall present a more detailed analysis of the properties of 
DB1 originating from the bottom optical mode $k=45$. This breather is 
localized on the first turn,  from site 10 to site 14, as it is shown in
Fig.~\ref{modo45_t0.3}.  The energy redistribution dynamics within the DB can be appreciated 
from Fig.~\ref{energiaDBmodo45}~(a).
In particular, it is clear that the two site pairs (12,13) and (11,14) 
exhibit approximately phase-locked oscillations, while they are
anti-phase locked among them.
This is another illustration of the fact that the DB exchanges little or no 
energy with the surrounding.\\

%

It is instructive to analyse the different contributions to the total energies 
when the DB is present, in order to quantify to what extent the different
degrees of freedom  participate to the breather dynamics.  As it can be seen
from  Fig.~\ref{energiaDBmodo45}~(b), at one of the most energetic sites, $i=12$,  the
DB vibration involves essentially the backbone bonds ($\approx 84-86\, \%$) with
a  small contribution coming from the angular terms ($\approx 13-15\%$), while
the dihedral and Lennard-Jones  contributions are negligible. The observed 
ratios  between the different energetic contributions do not change by varying
the  initial energy. We observe the same behaviour for all the sites in the interval ($11-13$).\\

A closer inspection of Fig.~\ref{energiaDBmodo45}~(b) shows that 
the harmonic bond component oscillates at frequencies $2f_{DB}$
but also has a sizeable component at $f_{DB}$. 
This is due to the soft nonlinearity which is known to induce
a DC component in the displacement patterns of localized
modes~\cite{Bickham1993,Flach:2008vx}. To understand this in a 
simple way, let us consider two bond lengths
along the backbone at site $n$ , $r_1 \equiv r_{n-1,n}$ and $r_2 \equiv
r_{n,n+1}$. We may write their evolution as an oscillating part 
plus a static distortion of the bond lengths, 
\begin{eqnarray}
r_1(t)=A_1 \sin (\omega t)+\langle r_1 \rangle \nonumber\\
r_2(t)=A_2 \sin (\omega t)+\langle r_2 \rangle \nonumber
\end{eqnarray}
where $\omega=2\pi f_{DB}$ and the angular brackets represent a 
time average. Using the definition reported in Appendix A, it can 
be shown that the contribution of the two bonds 
to the local harmonic energy is 
\begin{eqnarray} 
\label{e:astuz}
V_{\rm H}(n) \propto  \frac{(A_1^{2}+A_2^{2})}{2} &\sin&^{2}(\omega t) \nonumber \\ 
&+&  [A_1(\langle r_1 \rangle-r_{0})+ A_2(\langle r_2 \rangle-r_{0})] \sin (\omega t) \\
&+& \frac{(\langle r_1 \rangle-r_0)^2+(\langle r_2 \rangle-r_0)^2}{2} \nonumber
\end{eqnarray}
Thus, the presence of a component at frequency $\omega$ implies that
$\langle r_1 \rangle$ and/or 
$\langle r_2 \rangle$ must be different from their equilibrium value 
$r_0$. Accordingly, we conclude that the emergence  of a DB also causes a stable 
structural distortion of the protein. 
Given that $A_1 \approx A_2 \approx 0.1$ (see again Fig.~\ref{energiaDBmodo45}), 
also tiny differences $\langle r_i \rangle -r_{0} \approx 10^{-3}$, 
as we recorded in our simulations, are able to affect the local bond energies.
      
\begin{figure}[ht!]
\begin{center} 
\includegraphics[width=0.6\textwidth,clip]{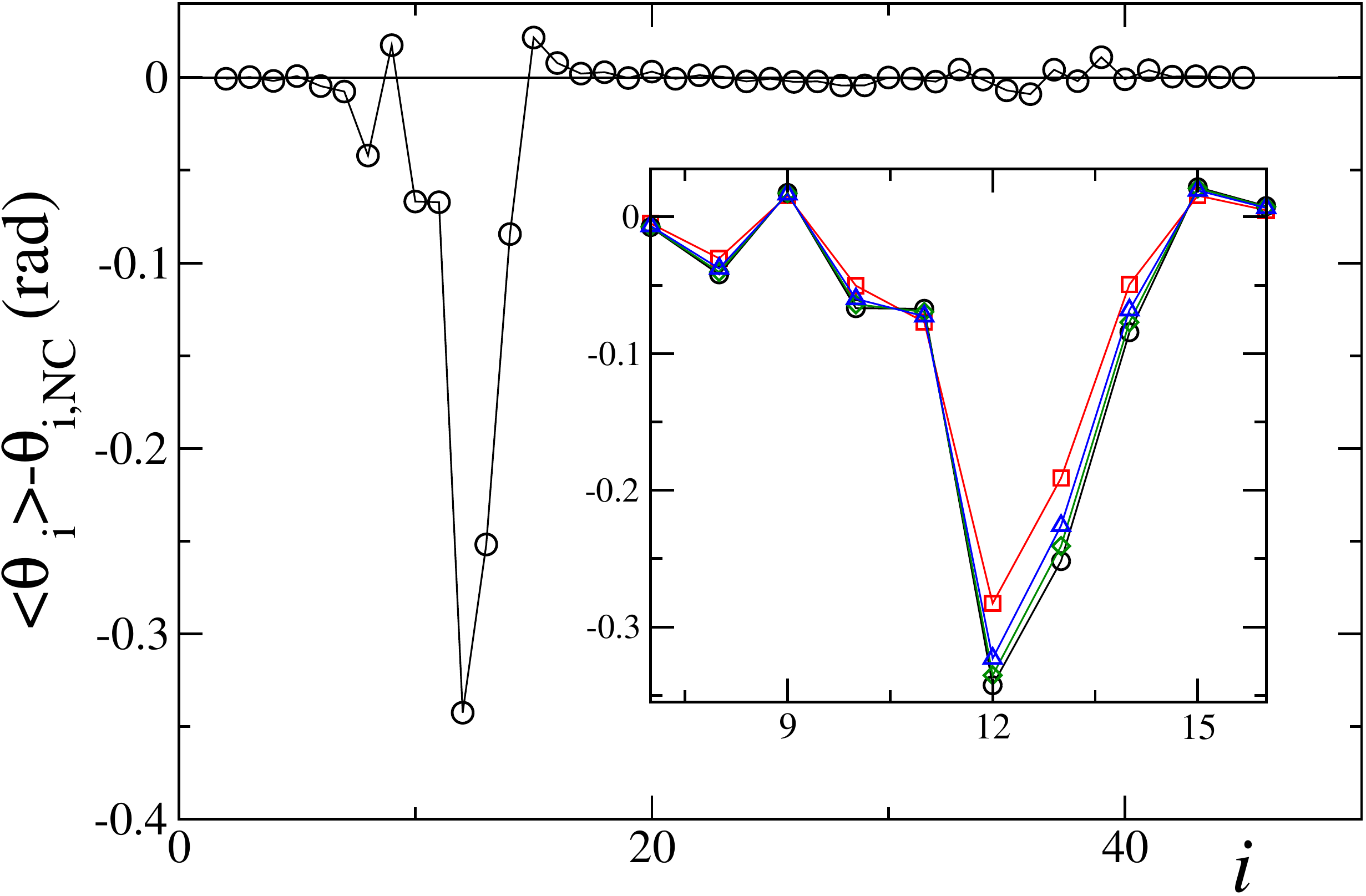}
\end{center}
\caption{(Color online) Difference between the average values 
of the bond-bond angles $\theta_{i}$ when the DB1 is present and the 
corresponding equilibrium values (in the NC), $\theta_{i,NC}$, when 
the total energy of DB1 is $E_{\scs DB}=34.6$. The inset shows a close-up of the kink region
for increasing energies:  (red) squares, (blue) triangles, (green) diamonds, 
(black) circles refer to $E_{\scs DB}=26.3,31.68,33.61,34.6$, respectively.}
\label{kink_modo45}
\end{figure}
%
%

Much more prominent is the distortion effect caused by the soft-nonlinearity 
in the angular degrees of freedom. In particular, 
Fig.~\ref{kink_modo45} reveals that the breather is also characterized 
by a kink-shaped angular distortion.  The kink's amplitude 
increases with the total energy stored in the breather 
(see the inset in Fig.~\ref{kink_modo45}). 

\subsubsection{DB collapse and the effect of cooling}
\label{nocooling}

\begin{figure}[b!]
\begin{center} 
\includegraphics[width=0.6\textwidth,clip]{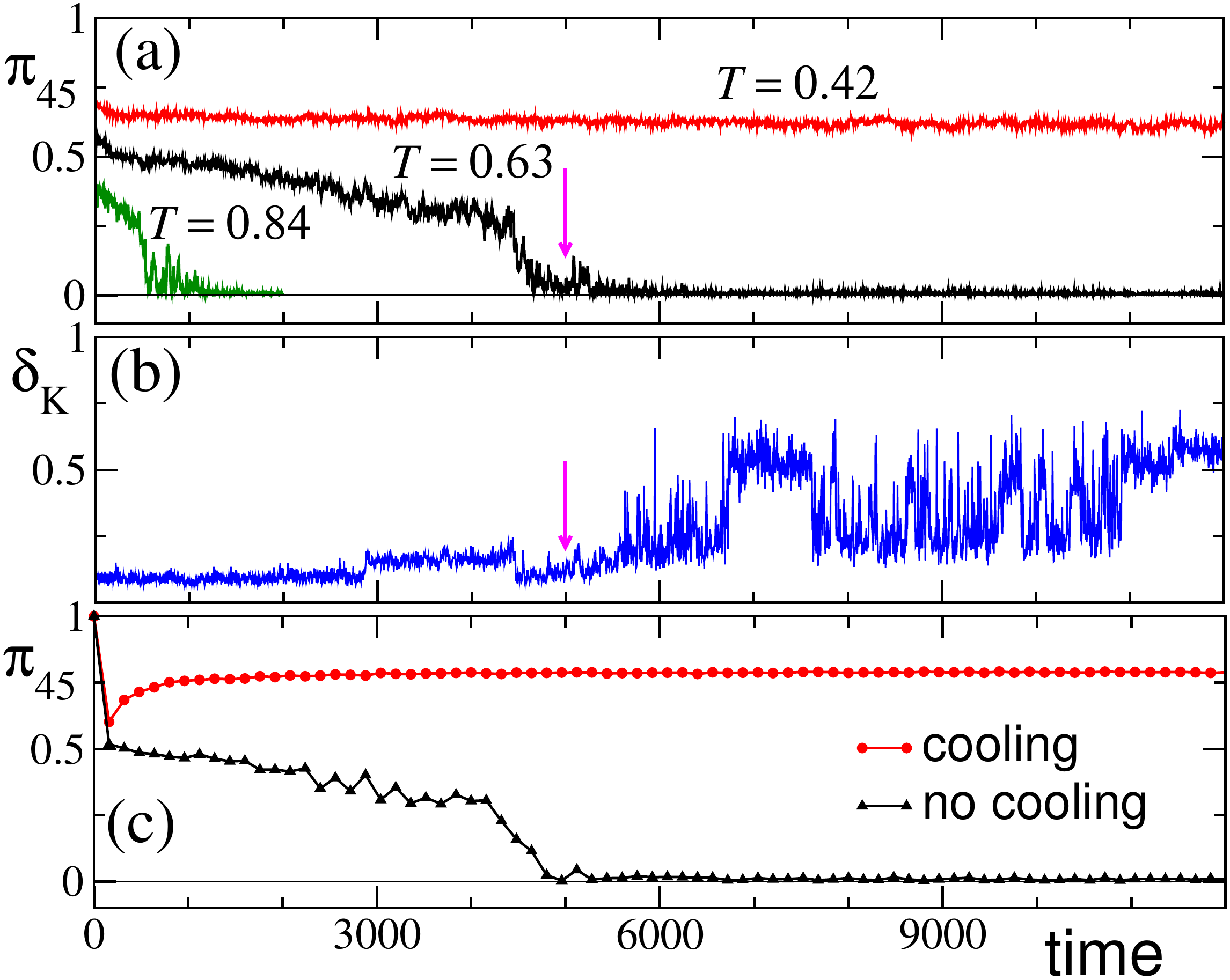}
\end{center}
\caption{(Color online) (a) Projection of the DB1 velocity field on the $k=45$ NM,
$\pi_{45}$, as a function of time 
for increasing values of the initial energy. From the top to the bottom curve,
$T=0.42,0.63,0.84$ (non-dissipative dynamics).
(b) Kabsch distance $\delta_{K}$ for $T=0.63$ (non-dissipative dynamics).
The vertical arrows mark the time at which the DB collapses.
(c) Comparison of  $\pi_{45}$ for dissipative (upper red line), 
and non-dissipative (lower black line) dynamics for $T=0.63$.
}
\label{cfr_fric_nofric}
\end{figure}

In the case where no cooling is applied ($\gamma=0$) the 
DBs have a finite lifetime. To illustrate this,
in Fig.~\ref{cfr_fric_nofric}~(a) we report the time
evolution of the projection  $\pi_{45}$ of the DB1,
obtained as a  continuation of the lower optical-band edge mode ($k=45$). 
The sudden drop of $\pi_{45}$ signals the collapse of the excitation,
accompanied by a rapid redistribution of energy over all the normal 
modes (equipartition). The typical collapse time increases upon decreasing
the initial energy (compare the three curves in Fig.~\ref{cfr_fric_nofric}~(a)).
This is similar to what demonstrated for one-dimensional chains
where the time to equipartition scales as an inverse power of the 
energy density~\cite{Cretegny1998,Kosevich2000}. 

The effect of the collapse of a DB on the protein structure can be  appreciated 
by considering the Kabsch distance $\delta_{K}$, which is a commonly used measure of the 
structural distance between two protein configurations~\cite{kabsch}.
In particular, Fig.~\ref{cfr_fric_nofric}~(b) shows how the 
distance $\delta_{K}$ of the protein structure from the NC evolves in time
in the presence of a discrete breather of type DB1.  As long as the DB is present, 
$\delta_{K}$ fluctuates around a
relatively small value ($\delta_K \simeq 0.1$), meaning that the protein structure does not deviate
appreciably from the NC. In fact, the static distortion effect illustrated above
only concerns a small region of the fold. On the other hand, when the DB collapses,
$\delta_{K}$ starts to fluctuate wildly around a  substantially larger value 
($\delta_K \approx 0.35$), thus signaling the occurrence of important conformational 
rearrangements. 

The DB lifetime gets substantially increased by removing the background vibrations through
cooling. Fig.~\ref{cfr_fric_nofric}~(c) clearly illustrates how
dissipation renders the DB stable by quickly eliminating background radiation.
The stabilizing role of the boundary cooling method is well 
documented~\cite{Aubry:96,Piazza:03,Juanico:07}
and our results confirm that it can be used conveniently also for 
such a complex structure.

\subsection{Excitation of optical band-edge NMs: long-range energy transfer}

\begin{figure}[ht!]
\begin{center} 
\subfigure[]{
\includegraphics[width=0.5\textwidth,clip]{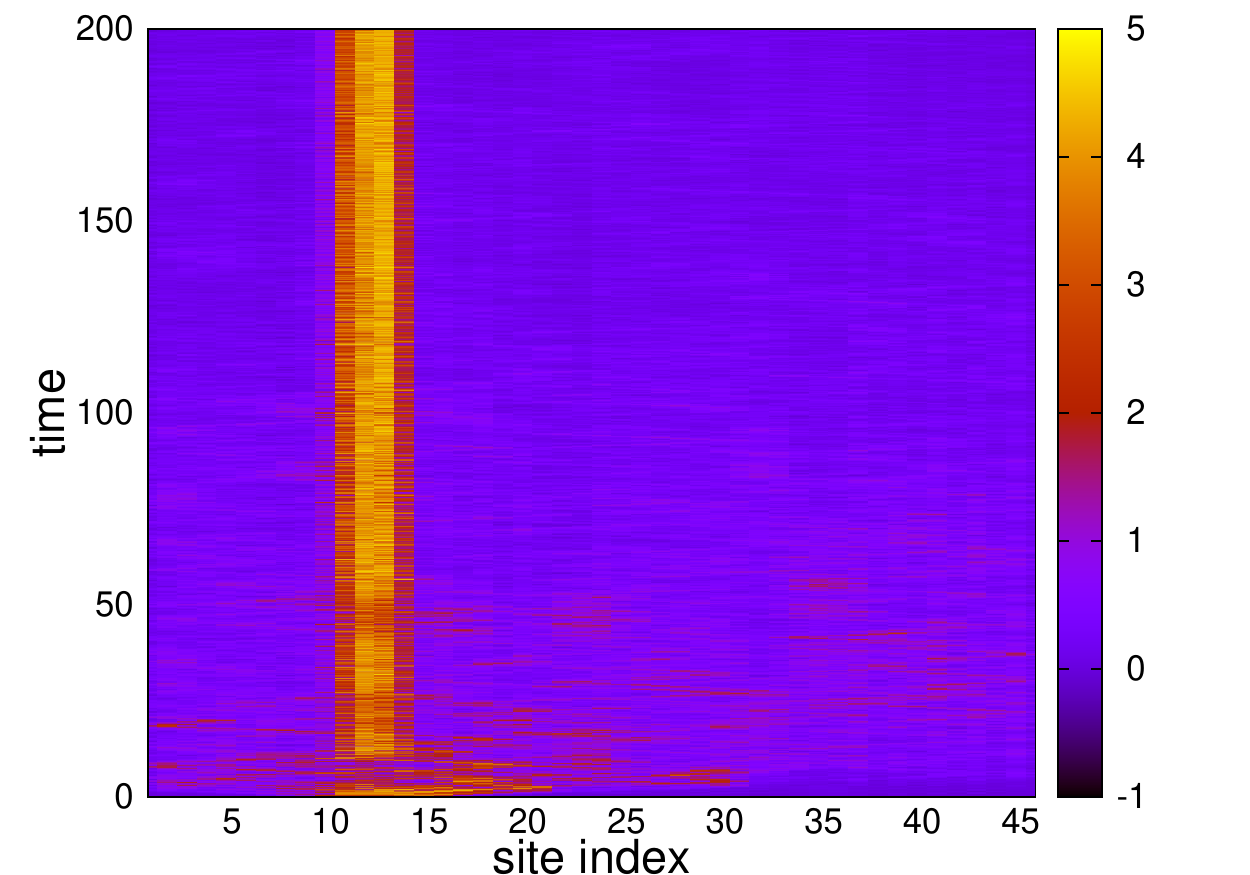}
}
\subfigure[]{
\includegraphics[width=0.5\textwidth,clip]{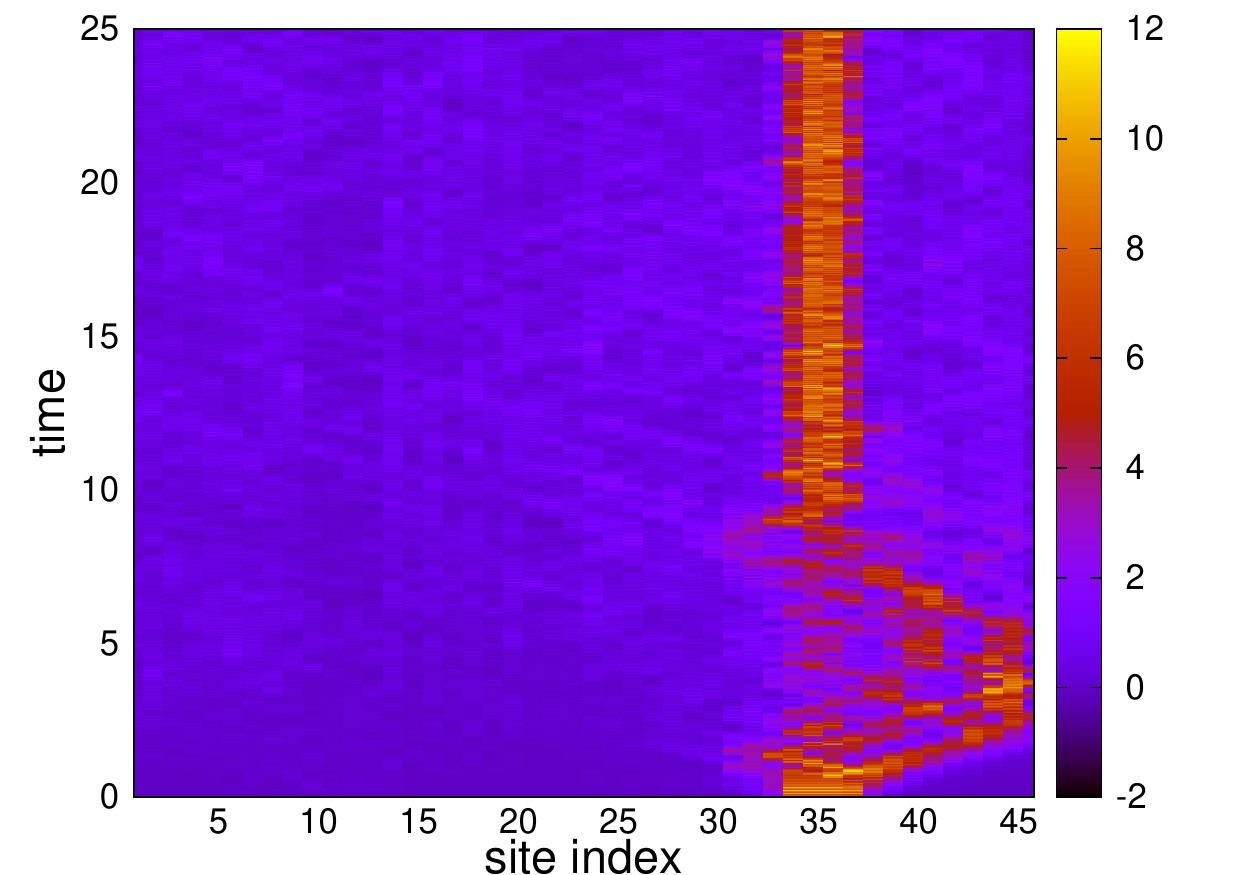}
}

\end{center}
\caption{(Color online) Time evolution of total site energies starting from  
excitation of the upper band-edge optical modes: 
(a) Emergence of DB1 from $\vec e_1$, $T=0.63$ and 
(b) of DB3 from $\vec e_2$, $T=0.84$.
}
\label{modi2-3}
\end{figure}

The excitation of NMs at the upper edge of the spectrum gives rise to quite
peculiar phenomena. As a first example, Figs.~\ref{modi2-3} 
depict the emergence of a DB obtained by exciting the first two 
highest-frequency NMs, i.e. $\vec e_1$, which is localized at sites $10-12$ 
and $\vec e_2$, which is localized at sites $34-36$. In the
first stage,  the energy quickly spreads and remains confined in the vicinity of the
perturbed location. After some time, a DB self-localizes, harvesting energy
from the background and pinning it in the same region. The DBs obtained are again 
DB1 and DB3. This is reasonable in view of the similar spatial structure 
of $\vec e_1$, $\vec e_2$ and DB1, DB3, respectively 
(see again Fig.\ref{modinormali}). Moreover, we found that 
the frequencies of DBs originating from $\vec e_1$ lie on the same  
frequency-energy curve as the breather originating from normal mode 45, 
shown in Fig.~\ref{frequenze_DB}.\\


\begin{figure}[ht!]
\begin{center} 
\subfigure[]{
\includegraphics[width=0.5\textwidth,clip]{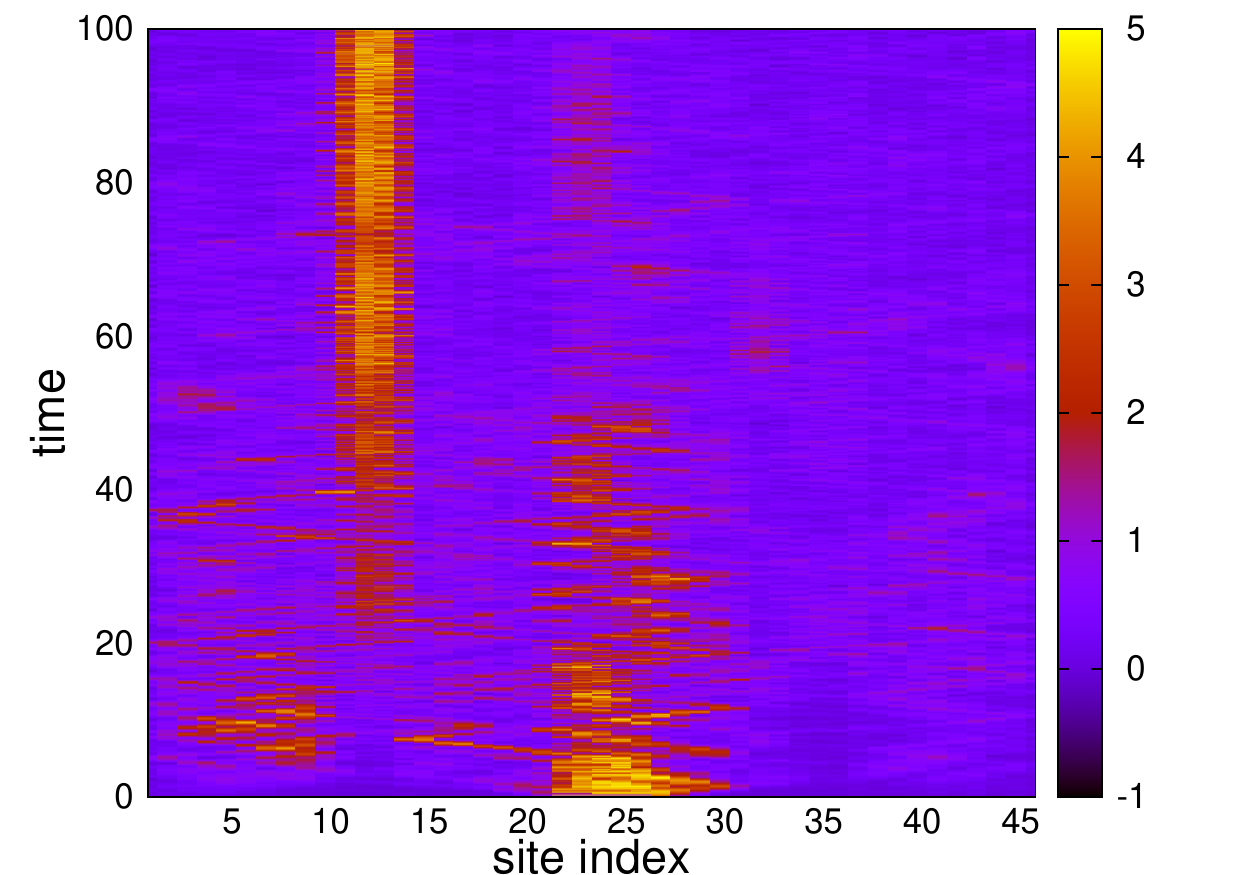}
}
\subfigure[]{
\includegraphics[width=0.45\textwidth,clip]{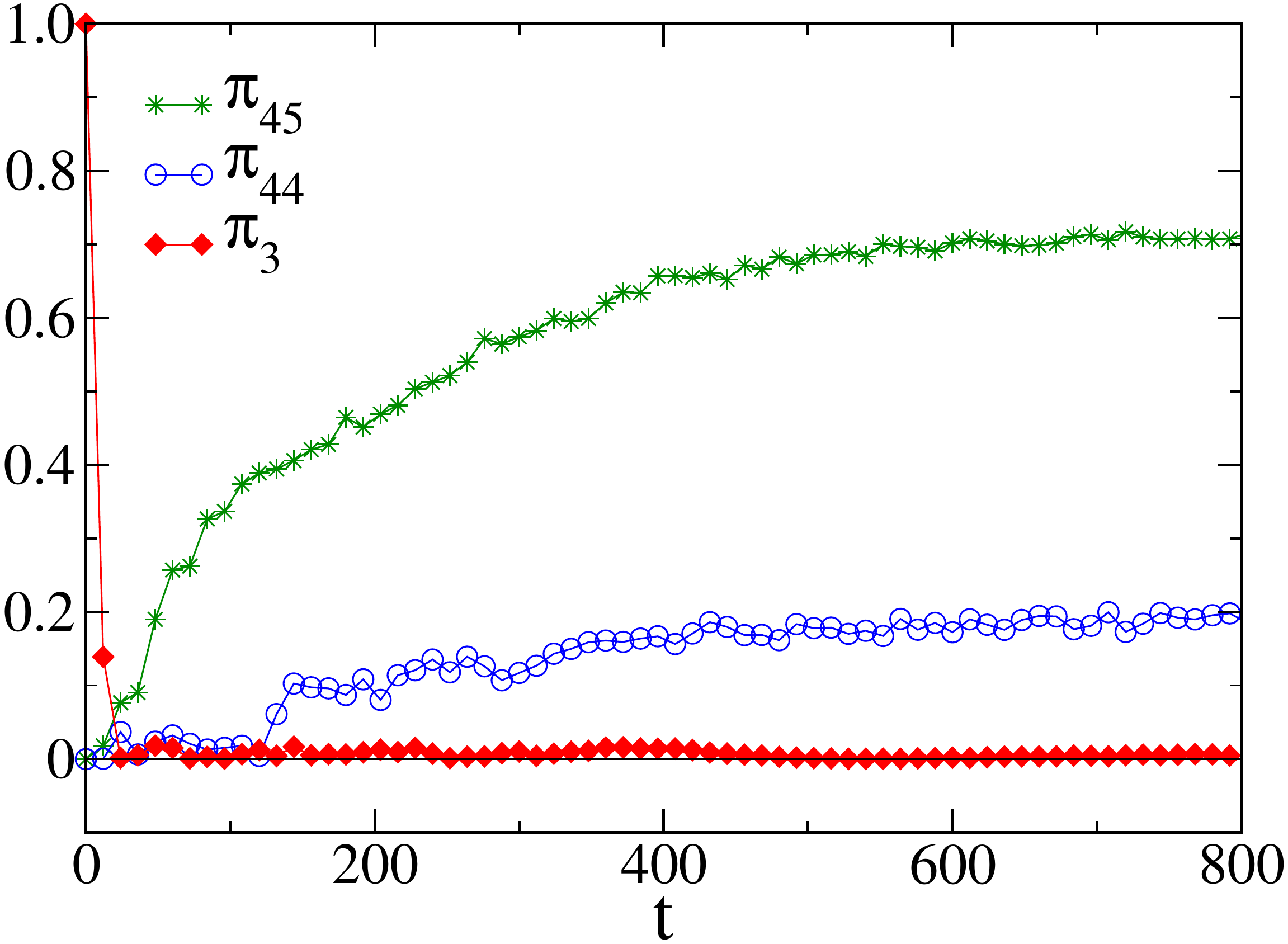}
}
\end{center}
\caption{(Color online) Long-range energy transfer starting 
from $\vec e_3$ at $T=0.63$. (a) Time-evolution of total site energies; 
(b) projections of the DB velocity field on three NMs 
as functions of the time. 
}
\label{proiezioni_modo3}
\end{figure}

Remarkably, we have also observed that the excitation of an edge mode may
result in a \textit{long-range energy transfer} event. An illustration of this
phenomenon is given in Fig.~\ref{proiezioni_modo3}(a), for the excitation of the third
highest-frequency NM. The NM $\vec{e_{3}}$ is localized at sites $22-27$.
Immediately after the perturbation,  the DB appears localized in the same
region, but seems to collapse rapidly afterwards, spreading  its energy evenly
across the structure. However, after a considerable time span, another DB
emerges at a different location, namely at sites $11-13$. This is the region
where the lower optical-band edge mode ($k=45$) is localized. A substantial
part of the initial energy has been transferred and pinned down irreversibly
at the other end of the structure, covering the distance from a turn to the
following one (see again the cartoon depicting the protein structure).

Such energy transfer phenomenon can be rationalized by analyzing the 
projections of the DB velocity field on the NMs during the time  evolution of
the perturbation.  From  Fig.~\ref{proiezioni_modo3}~(b) one can clearly see
that  the initially excited mode is quickly emptied of its initial energy,
which gradually flows into other two modes. As it shows, asymptotically  the
two target modes almost describe the entirety ($\pi_{44} + \pi_{45} \approx 1$)
of the DB, which is in fact localized at the  turn opposite to the excited one,
where the two target NMs are also localized.
The reverse process is forbidden since the DB frequency is no longer 
resonating with the original mode. We remark that this phenomenon, although evocative of 
resonant energy transfer among few selected normal modes in proteins,
is a one-way transfer, as the DB will retain the transferred energy for 
times comparable to its lifetime~\cite{K:2003hq,K:2000kh}.

\subsection{Excitation of deep-band NMs}

\begin{figure}[ht!]
\begin{center} 
\subfigure[]{
\includegraphics[width=0.5\textwidth,clip]{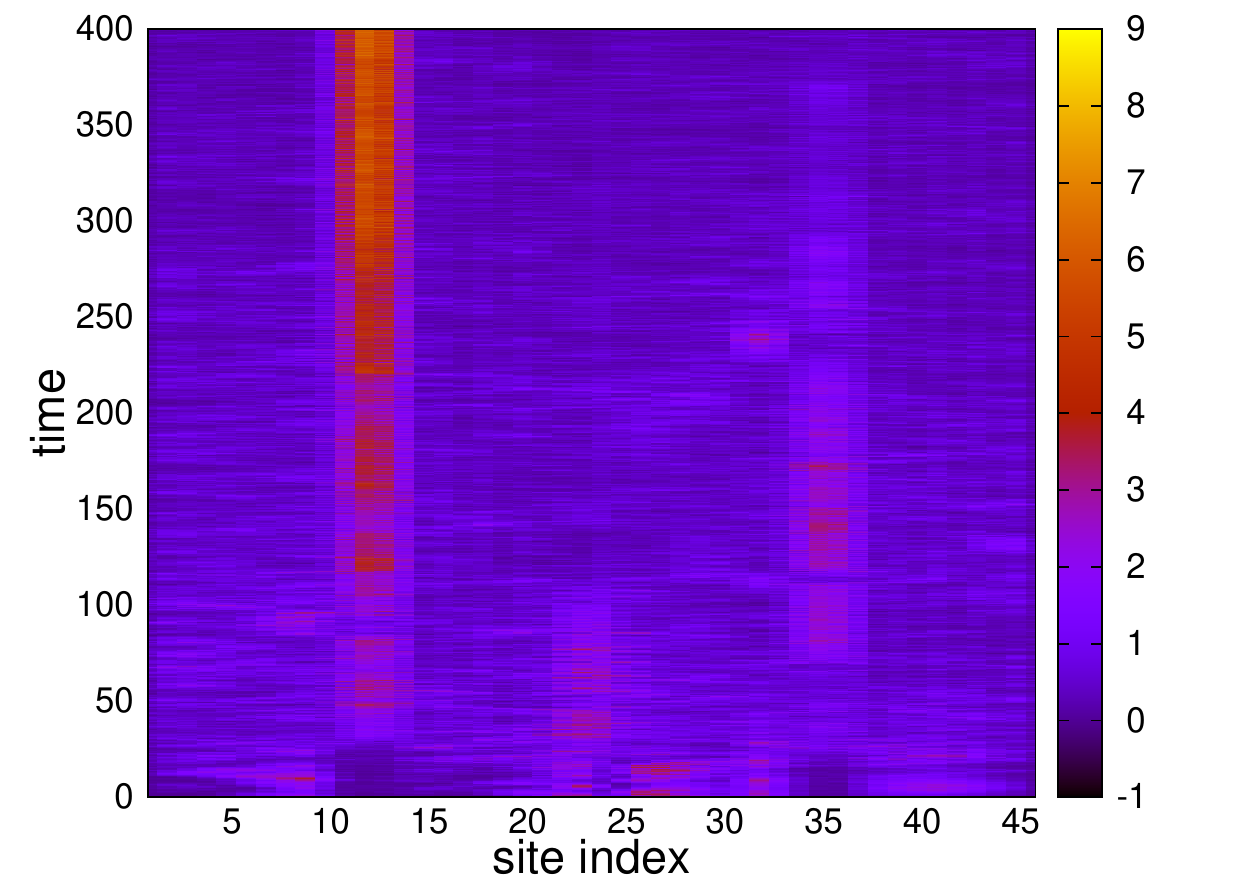}
}
\subfigure[]{
\includegraphics[width=0.5\textwidth,clip]{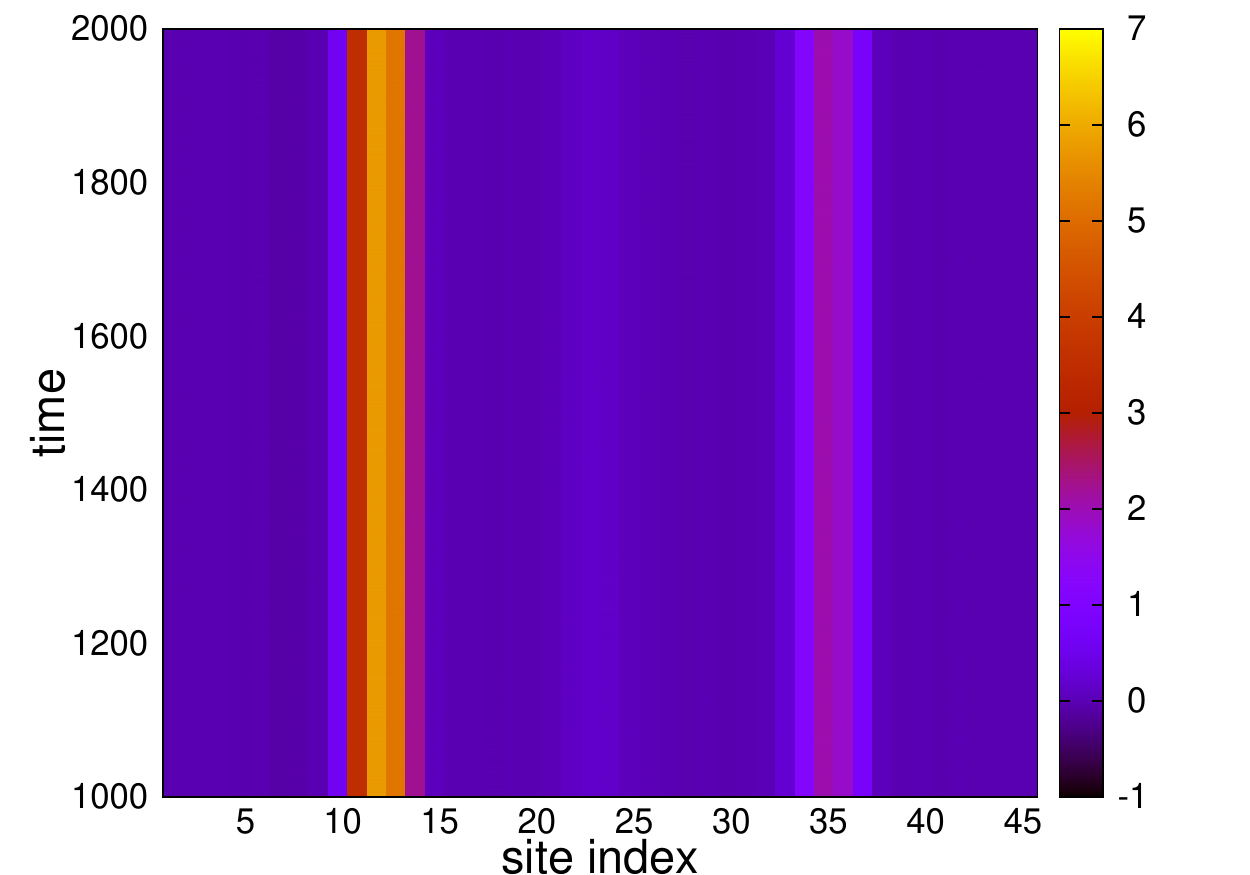}
}
\subfigure[]{
\includegraphics[width=0.5\textwidth,clip]{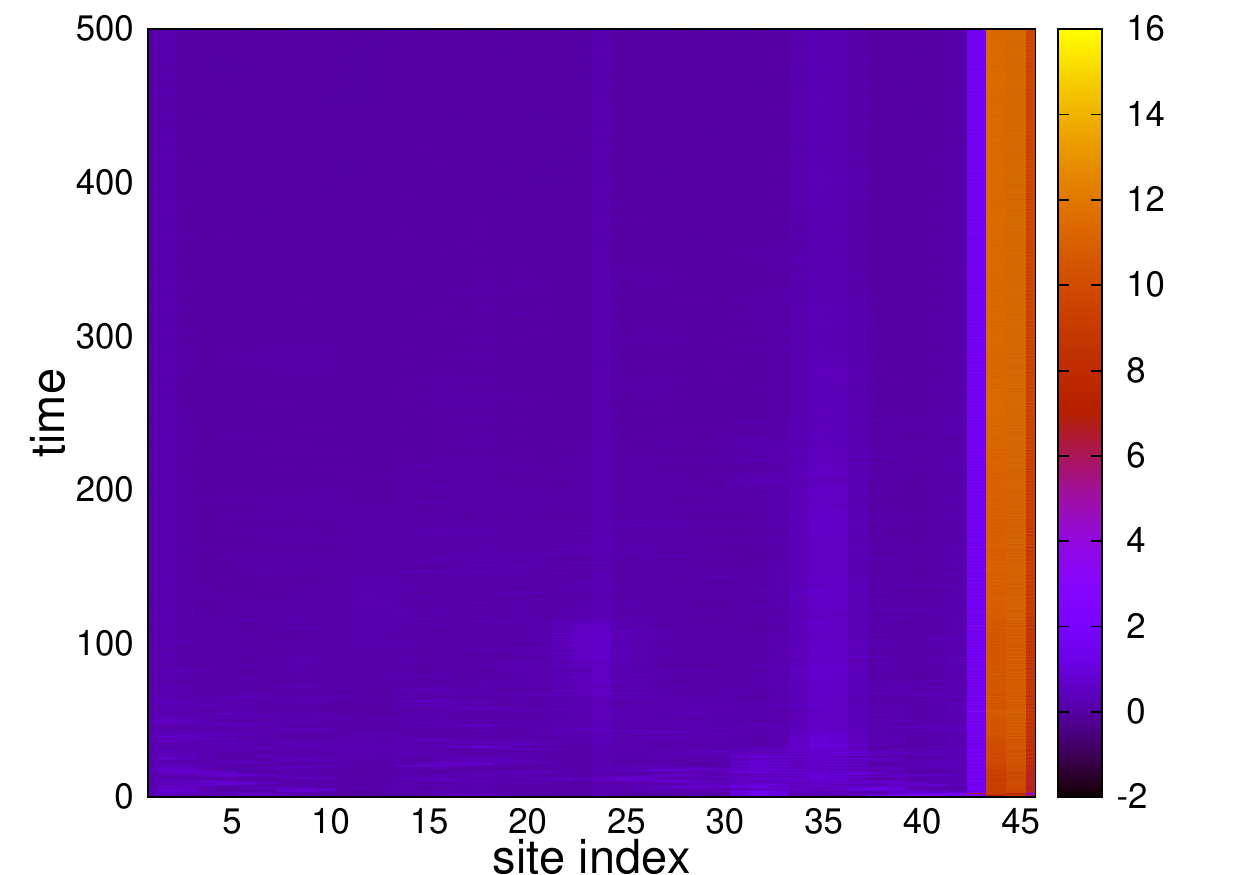}
}
\end{center}
\caption{(Color online) Evolution of total site energies starting from  
excitation of the deep-band optical modes: 
(a) Emergence of DB1 from $\vec e_5$, $T=0.63$ and 
(b) multibreather solution localized on the first and third turn 
starting from $\vec e_{33}$ at $T=0.63$; (c) 
Boundary Breather (BB) following the excitation 
of mode $\vec e_{36}$, $T=0.84$ (non-dissipative dynamics).
}
\label{modi5-36}
\end{figure}

NMs in the core of the optical band are generally characterized by a  low level of
localization, as one can see in Fig.~\ref{figuravarioalpha} (b-e). As a result,
when one of such modes is excited, we observe a complex localization  pattern, that
alternates site-hopping and energy-pinning stages to phases where the  energy is
more evenly distributed.  After a certain time, a stable DB emerges, focusing the
energy at one of the turns. A typical realization of this scenario is illustrated
in Fig.~\ref{modi5-36}(a) for  the excitation of the fifth highest-frequency mode.

A different phenomenology is observed when we excite a mode lying deeper within the 
band. It may occur that 
the excitation energy remains trapped in two DBs localized on two different turns, 
realizing a state which is reminiscent of multibreather states observed in 
non-linear disordered systems~\cite{Kopidakis:2000lr,Kopidakis:1999fk}. 
A realization of this phenomenon for NM 33 is reported in Fig.~\ref{modi5-36}~(b). 
In particular, we notice that starting from NM 33 and changing the initial energy 
we can get different solutions. For example, for $T=0.21$ we obtain a solution localized on the 
first and second turn, while the multibreather localizes on the first and third ones for $T=0.63$.

\begin{figure}[ht]
\begin{center} 
\includegraphics[width=0.7\textwidth,clip]{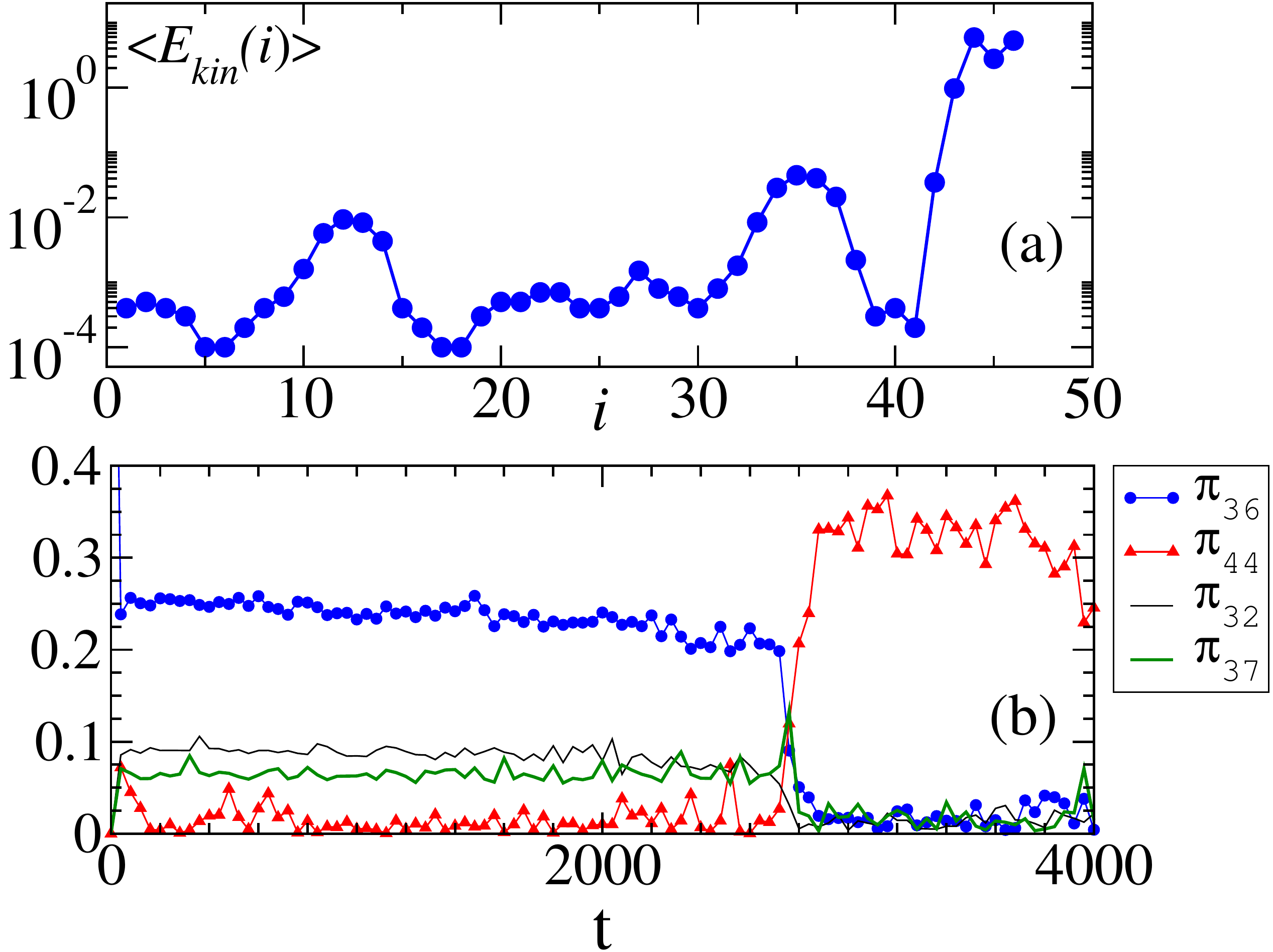}
\end{center}
\caption{(Color online) Emergence of a Boundary Breather (BB) following the excitation 
of mode $k=36$ with $T=0.63$. (a) time-averaged 
site kinetic energies (the cooling is applied to the first 4 beads); 
(b) projections of the DB normalized velocity field on the NMs as a 
function of time (non-dissipative dynamics).}
\label{boundarymode_36}
\end{figure}

A very peculiar localized solution develops through the excitation of normal 
mode $k=36$, whose pattern  is localized in the protein's tail.  As it is shown
in Fig.~\ref{modi5-36}~(c), the energy remains almost entirely confined to 
the edge sites $44-46$, with smaller, but  non-negligible, amounts of energy also
involving the turns Fig.~\ref{boundarymode_36}~(a).  We term such excitations 
{\em boundary breathers} (BB). 
As in the case of DB1, DB2 and DB3, the frequency of oscillation of the BB
also lies in the gap ($f_{BB}=7.739$ for $E_{\scs DB}=34.75$). 
At variance with DB1, DB2, DB3, the energy of  BBs has not only harmonic and
angular contributions but also a sizeable Lennard-Jones one, 
resulting from a non-negligeable interaction between the first and last beads, 
as well as the second turn region of the chain
which are relatively close in the NC. 
We note that nonlinear surface states of different kinds are known 
and studied in many contexts~\cite{Flach:2008vx,Lazarides2008}. 

\begin{figure}[ht!]
\begin{center} 
\subfigure[]{
\includegraphics[width=0.5\textwidth,clip]{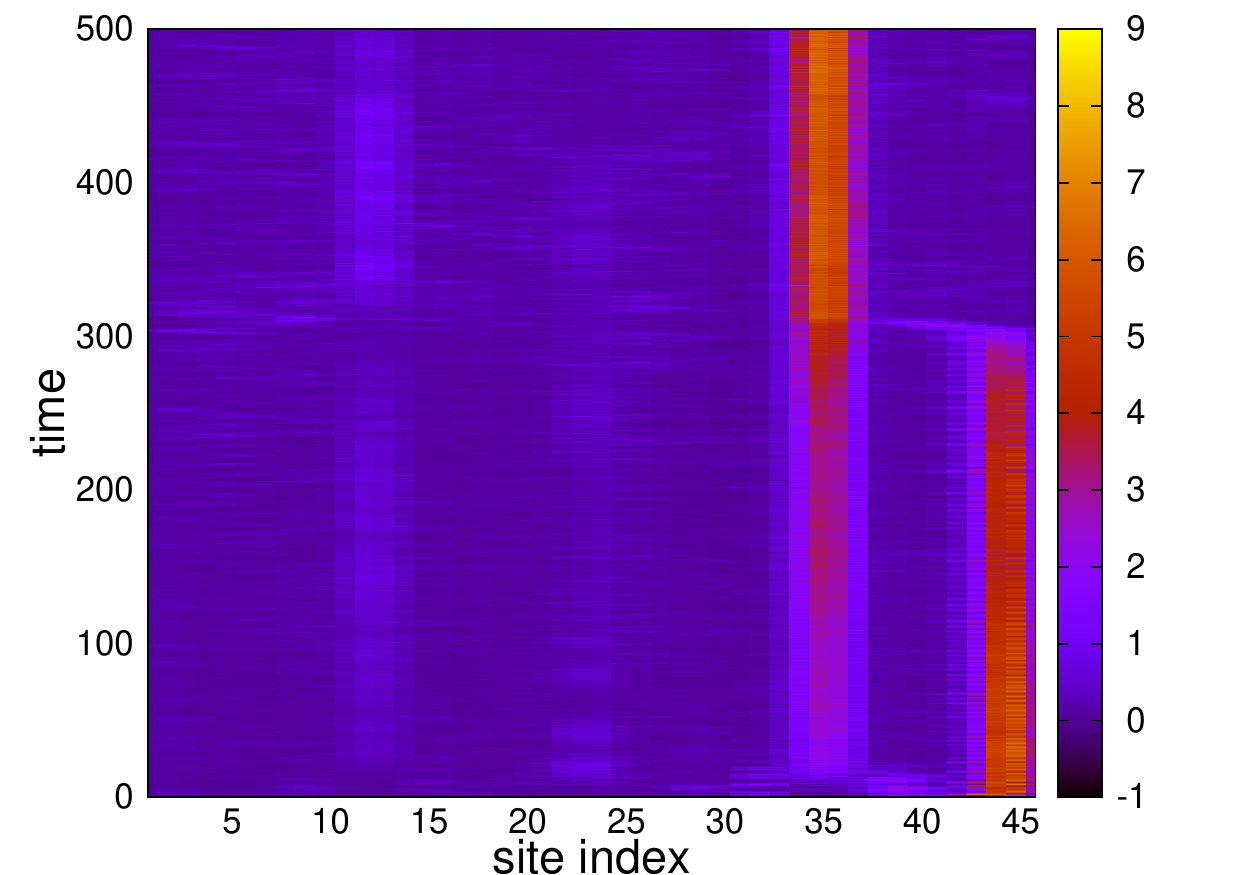}
}
\subfigure[]{
\includegraphics[width=0.45\textwidth,clip]{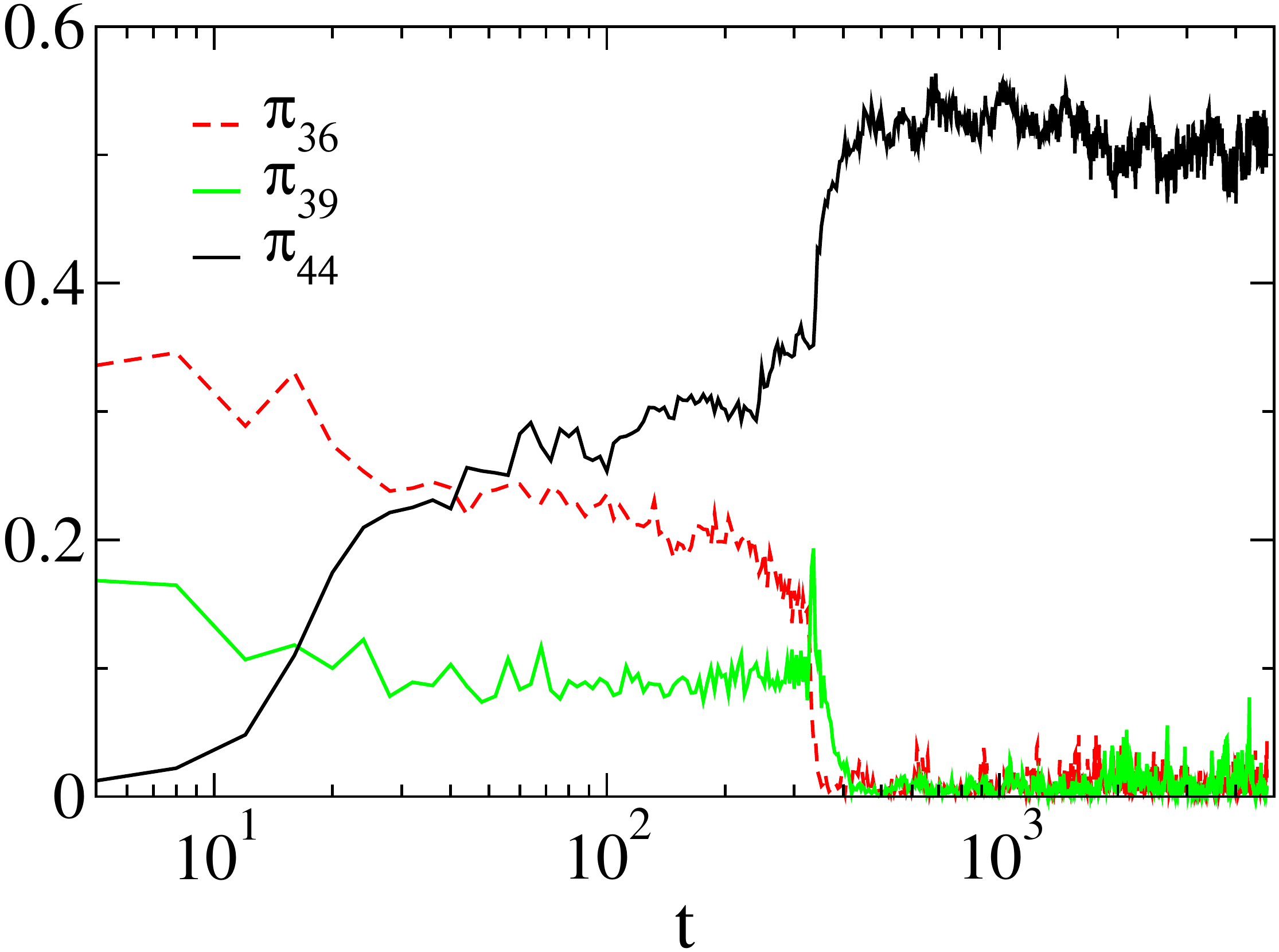}
}
\end{center}
\caption{(Color online) Transient multibreather-like mode, localized
on the protein boundary and on the third turn, obtained following the excitation 
of the $k=36$ NM at  $T=0.42$ (non-dissipative dynamics). 
(a) Space-time plot of the total site energies. 
(b) Normalized projections of the DB velocity field on the NMs 
as a function of the time.}
\label{bm_36_t0.2}
\end{figure}

Eventually, the BBs hop to one of the three breathers described above. 
Fig.~\ref{bm_36_t0.2} illustrates an example of this process, where the BB
transfers its energy to DB3, as confirmed
by the time evolution of the normal-mode projections
shown in Fig.~\ref{boundarymode_36}~(b).  Asymptotically,
we see that the main projection is $\pi_{36}$, while  at $t \approx 2.75 \times
10^3$ an abrupt transition occurs and the leading projection  becomes
$\pi_{44}$, signalling the energy transfer to DB3 (corresponding to NM $k=44$)
localized at the third turn. 

While at high energies, {\em i.e.} $T=0.84,1.05$, the lifetime of BBs before 
a transition to DB3 takes place is short, at lower energies 
we observe the formation of a transient 
multibreather-like state with a progressive transfer of energy from BB to DB3.
One of such examples is depicted in Fig.~\ref{bm_36_t0.2}. Decreasing the excitation
energy further, BBs are no longer observable.

We note that the end of the protein can be considered as a  defect/discontinuity 
in the chain, since it is a free end that interrupts the chain itself.
Therefore, BBs can be considered as akin to localized modes
emerging in other contexts close to point defects.
Furthermore,  we were unable to excite stable BBs
localized on the first beads of the sequence, despite  NMs 
with significative components on that terminal do exist. A possible
explanation of this apparently contradictory result is the fact
that the first strand is part of the protein core. Therefore, it is
a much more rigid structure with respect to the last strand,
which instead is exposed to the solvent and can oscillate more freely.

\section{Conclusions}

In this paper we have reported the existence of discrete breathers
in an off-lattice model of proteins with realistic interaction potentials and 
three-code amino-acid sequence. 
To our knowledge, this is the first time that this class of peculiar vibrational modes 
are found in such a complex and heterogeneous dynamical system.
In particular, we have shown that, due to the softness of the leading 
nonlinearity, a necessary condition for DB existence
is that the linear spectrum exhibits a gap (at least one), as it is indeed the case in all 
proteins. 
%
%
%

For the particular structure  analyzed in this work, we have identified three families of DBs, 
each one localized on a different turn of the native fold. The 
largest fraction of the energy of a discrete breather
typically resides on a few sites and involves essentially  harmonic and angular degrees 
of freedom, while dihedral and Lennard-Jones interactions between 
distant sites are practically not excited. 

We have obtained discrete breathers as continuations of the lower optical 
band-edge normal modes. Quite generically, however, we have shown that members of
the three different breather families can be excited by feeding energy to 
{\em any} of the normal modes lying in the optical band.  
Remarkably, the excitation of a DB obtained by feeding energy to a 
given normal mode is an extremely {\em efficient} process. This means that 
there is an extended range of initial energies that can be fed to the system and 
immediately channeled almost entirely into a breather, which is permanently
localized at a very specific location.

%
%

Our results prove that  DBs have a stabilizing effect on the protein. 
In fact, a broad class of perturbations of the native fold 
result in the formation of a DB, that is, a stable and highly localized 
vibration, while on the other hand long-range contacts are almost devoid of energy. 
This enhances the structural  stability of the protein. To this regard, it 
would be interesting to connect the structure-dynamics relation underlying 
DB formation to the well-conserved  patterns of nucleation sites along 
folding pathways~\cite{Mirny2001123}.

We also discovered peculiar DB-assisted long-range energy transfer phenomena,
whereby  energy is channeled to a distant region of the structure away from the excitation site. 
At variance with ordinary energy exchange between resonant normal 
modes~\cite{Kidera:00}, this is a one-way process, meaning that the 
transferred energy is never released back to the starting location.
In particular, this effect may be of importance in the functioning of 
allosteric proteins~\cite{Tsai:2009ad,Sol:2009zp,GunMaNus04} 
and surely deserves  further investigation. 
           
Our work has highlighted nontrivial correlations between structural and 
dynamical features of protein folds. In a simple $\beta$-barrel, as the one
here considered, localization occurs on loops. Concerning more complex structures,
our results raise the important question whether more complex structural selection rules 
exist driving nonlinear energy localization and long-range targeted transfer.
 

\ack
We acknowledge useful discussions with S. Flach, R. Livi and  L. Schimansky-Geier.
Thanks are due to S. Allaccia and A. Scarpa for inspiring the 
style and presentation of the paper. 
AT gratefully acknowledges the financial support of 
Aarhus Universitet Forskning Fonden (AUFF) during his stay at Aarhus University.
AI is partially supported by Lundbeck Fonden.
This work is part
of the CNR-RSTL project N. 827 \textit{Dinamiche cooperative in
strutture  quasi uni-dimensionali}.
Finally, we thank the Danish Centre for Scientific Computing for providing
us with computational resources.

\appendix\section*{Appendix A: Single Site Energy}

Since each potential energy term represents a many-body interaction,
we have estimated the single-site potential energies $V(i)$ by equally redistributing
the energy to each site involved in the different terms and then by
summing up all this contributions. 

In particular, the harmonic term is a two body potential involving
nearby sites, therefore the contribution is estimated as follows
$$
V_{H}(i)=\left\{ \begin{array}{lll}
\vspace{0.1cm}
\sum_{k=i}^{i+1}V_{H}(r_{k-1,k})/2    & \mbox{for} \quad i=2,L-1     \\
\vspace{0.1cm}
V_{H}(r_{1,2})/2    & \mbox{for} \quad i=1 \\
\vspace{0.1cm}
V_{H}(r_{L-1,L})/2    & \mbox{for} \quad i=L
\end{array} \right. 
\qquad .
$$

The angular term is a three body interaction involving three consecutive
sites, and therefore
$$
V_{A}(i)=\left\{ \begin{array}{lllll}
\vspace{0.1cm}
\sum_{k=i-1}^{i+1}V_{A}(\theta_{k})/3     & \mbox{for} \quad i=3,L-2    \\
\vspace{0.1cm}
V_{A}(\theta_{2})/3    & \mbox{for} \quad i=1 \\
\vspace{0.1cm}
\sum_{k=2}^{3}V_{A}(\theta_{k})/3   & \mbox{for} \quad i=2\\
\vspace{0.1cm}
\sum_{k=L-2}^{L-1} V_{A}(\theta_{k})/3   & \mbox{for} \quad i=L-1\\
\vspace{0.1cm}
V_{A}(\theta_{L-1})/3  & \mbox{for} \quad i=L
\end{array} \right. 
\qquad .
$$

The dihedral angle $\varphi$ is the angle between two nearby planes
each containing three consecutive sites, the corresponding potential term
is therefore a four-body term, and the single site contribution can be evaluated as follows
$$
V_{D}(i)=\left\{ \begin{array}{lllllll}
\vspace{0.1cm}
\sum_{k=i-2}^{i+1}V_{D}(\varphi_k, \theta_k, \theta_{k+1})/4     & \mbox{for} \quad i=4,L-3     \\
\vspace{0.1cm}
V_{D}(\varphi_2, \theta_2, \theta_{3})/4    & \mbox{for} \quad i=1 \\
\vspace{0.1cm}
\sum_{k=2}^{3} V_{D}(\varphi_k, \theta_k, \theta_{k+1})/4   & \mbox{for} \quad i=2\\
\vspace{0.1cm}
\sum_{k=2}^{4} V_{D}(\varphi_k, \theta_k, \theta_{k+1})/4    & \mbox{for} \quad i=3\\
\vspace{0.1cm}
\sum_{k=L-4}^{L-2} V_{D}(\varphi_k, \theta_k, \theta_{k+1})/4   & \mbox{for} \quad i=L-2\\ 
\vspace{0.1cm}   
\sum_{k=L-3}^{L-2} V_{D}(\varphi_k, \theta_k, \theta_{k+1})/4   & \mbox{for} \quad i=L-1\\
\vspace{0.1cm}
V_{D}(\varphi_{L-2}, \theta_{L-2}, \theta_{L-1})/4   & \mbox{for} \quad i=L
\end{array} \right. 
\qquad .
$$

The Lennard-Jones term is a two-body interaction involving sites 
separated by more than two bonds along the chain
$$
V_{LJ}(i)=\sum_j V_{LJ}(r_{ij})/2  \hspace{0.2cm}  {\rm for} \quad |i-j| > 2  
\qquad .
$$
The total site potential energies read
$$
V(i) = V_H(i) + V_A(i) + V_D(i) + V_{LJ}(i).
$$

\section*{References}

\providecommand{\newblock}{}

\end{document}